\documentclass[numsec,webpdf,modern,medium,namedate]{oup-authoring-template}

\usepackage{amsthm}

\newtheoremstyle{tight}%
  {4pt}   
  {2.5pt}   
  {\itshape}{}{\bfseries}{.}{.5em}{}  

\theoremstyle{tight}

\usepackage{xr}
\usepackage{lmodern}
\onecolumn
\usepackage{bbm}

\allowdisplaybreaks[4]

\graphicspath{{Figures/}}

\newtheorem{theorem}{Theorem}
\newtheorem{lemma}{Lemma} 

\newtheorem{remark}{Remark}
\newtheorem{definition}{Definition}
\usepackage[doublespacing]{setspace}
\usepackage[fontsize=12pt]{fontsize}

\AtBeginDocument{%
    \setlength{\abovedisplayskip}{4pt plus 2pt minus 2pt}
    \setlength{\belowdisplayskip}{4pt plus 2pt minus 2pt}
    \setlength{\textfloatsep}{2pt plus 2pt minus 4pt}
    \setlength{\floatsep}{1.5pt plus 2pt minus 2pt}
    \setlength{\intextsep}{2pt plus 2pt minus 2pt}
    \setlength{\abovedisplayshortskip}{4pt plus 2pt minus 2pt}
    \setlength{\belowdisplayshortskip}{4pt plus 2pt minus 2pt}
}

\usepackage{etoolbox}  

\makeatletter
\patchcmd{\@maketitle}
  {\@abstract}%
  {\begingroup              
     \setlength{\baselineskip}{16pt}
     \@abstract
   \endgroup}%
  {}{}                      
\makeatother

\begin{document}

\journaltitle{} 
\DOI{DOI HERE}
\copyrightyear{XXXX}
\pubyear{XXXX}
\access{Advance Access Publication Date: Day Month Year}
\appnotes{Original article}

\firstpage{1}


\title[Inferable Private Multiple Testing]{SUP: An Inferable Private Multiple Testing Framework with Super Uniformity}

\author[1,2]{Kehan Wang}
\author[1]{Wenxuan Song}
\author[1,$\ast$]{Wangli Xu}
\author[2,$\ast$]{Linglong Kong}


\authormark{Wang et al.}

\address[1]{Center for Applied Statistics and School of Statistics, Renmin University of China, Beijing, 100872, China}
\address[2]{Department of Mathematical and Statistical Sciences, University of Alberta, Edmonton, AB, T6G 2G1, Canada}

\corresp[$\ast$]{Address for correspondence. Wangli Xu, Center for Applied Statistics and School of Statistics, Beijing, 100872, China. \href{Email:email-id.com}{wlxu@ruc.edu.cn}; Linglong Kong, Department of Mathematical and Statistical Sciences, University of Alberta, Edmonton, AB, T6G 2G1, Canada. \href{Email:email-id.com}{lkong@ualberta.ca}.}

\received{Date}{0}{Year}
\revised{Date}{0}{Year}
\accepted{Date}{0}{Year}




\abstract{Multiple testing is widely applied across scientific fields, particularly in genomic and health data analysis, where protecting sensitive personal information is imperative. However, developing private multiple testing algorithms for super uniform $p$-values remains an open question, as privacy mechanisms introduce intricate dependence among the peeled $p$-values and disrupt their super uniformity, complicating post-selection inference. To address this, we introduce a general Super Uniform Private (SUP) multiple testing framework with three key components. First, we develop a novel \( p \)-value transformation that is compatible with diverse privacy regimes while retaining the super uniformity. 
Next, a reversed peeling algorithm is designed to reduce privacy budgets while facilitating inference. Then, we provide diverse rejection thresholds that are privacy-parameter-free and tailored for different Type-I errors, including the family-wise error rate (FWER) and the false discovery rate (FDR). Building upon these, we advance adaptive techniques to determine the peeling number and boost thresholds. Theoretically, we propose a technique overcoming the post-selection obstacle to Type-I error control, quantify the privacy-induced power loss of SUP relative to its non-private counterpart, and demonstrate that SUP surpasses existing private methods in terms of power. The results of extensive simulations and a real data application validate our theories.}

\keywords{differential privacy, false discovery rate, family-wise error rate, multiple testing, post-selection inference}


\maketitle

\section{Introduction}

Advances in data collection and biotechnology have enabled access to millions of personal genome sequences \citep{shi2017overview}, offering unprecedented opportunities to deepen our understanding of human biology and improve data-driven decision making. However, these benefits entail significant privacy risks, especially when managing sensitive information like genetic records.
Even minor allele frequencies can reveal individuals' participation in genome-wide association studies (GWAS), potentially exposing private disease statuses and causing personal harm or societal stigma \citep{homer2008resolving}. Such risks discourage data sharing, consequently hindering scientific progress. A case in point is that the U.S. National Institutes of Health strengthened data-access restrictions in their funded studies \citep{dwork2021differentially}. To address this, \citet{dwork2006calibrating} proposed differential privacy (DP), a rigorous framework that ensures the inclusion or exclusion of any individual’s data has a negligible impact on the analysis outcome, safeguarding confidentiality against external threats. DP has since become a cornerstone of privacy-preserving data analysis in fields such as bioinformatics, healthcare, and public administration.



Large-scale multiple testing frequently arises in genomic and human disease studies \citep{tian2003Erm, kaplanis2020evidence, rashkin2020pan, wingo2021integrating}, many of which involve sensitive data, aiming to detect signals while controlling Type-I errors across numerous hypotheses or variables. 
Key Type-I errors include the family-wise error rate \citep[FWER,][]{holm1979simple} and the false discovery rate \citep[FDR,][]{benjamini1995controlling}, defined respectively as
\begin{equation}
    \label{eq:Type-I error}
    \operatorname{FWER}=\mathbb{P}(V>0), \quad \operatorname{FDR} = \mathbb{E}\left[\frac{V}{R \vee 1}\right],
\end{equation}
where $R$ represents the number of rejected hypotheses and $V$ denotes the number of incorrectly rejected true null hypotheses. FWER is a stringent criterion, mandated by regulatory agencies in safety-critical fields such as drug clinical trials and genetic testing \citep{duggal2008establishing,zhu2020family,greenstreet2021multi}. 
Common approaches to controlling FWER include the Bonferroni method \citep{bonferroni1936teoria}, Holm's step-down method \citep{holm1979simple}, and resampling method \citep{romano2005exact}. In contrast, FDR is less restrictive and typically offers greater statistical power than FWER. FDR control techniques encompass Benjamini-Hochberg (BH) type methods \citep{benjamini1995controlling, benjamini2001control}, weighted $p$-value methods \citep{storey2002direct,li2019multiple}, among others. For controlling Type-I errors, the aforementioned methods typically assume that the null $p$-values either follow the uniform distribution $U(0,1)$ or satisfy a weaker super uniform condition, defined as $\mathbb{P}\left(p_j \leq t\right) \leq t$ for any $t \in[0,1]$. The super uniform condition is widely recognized and commonly used in multiple testing. Additionally, the mirror conservative condition is another commonly used condition for FDR control, defined as $\mathbb{P}(p_i \in [a_1, a_2]) \leq \mathbb{P}(p_i \in [1-a_2, 1-a_1])$ for any $ 0 \leq a_1 \leq a_2 \leq 0.5$. This concept originates from \citet{lei2018adapt} and is subsequently adopted by \citet{lei2020general, tian2021powerful, leung2022zap}. Both the super uniform and mirror conservative conditions can be satisfied simultaneously, as exemplified by the uniform distribution \(U(0,1)\). However, neither condition implies the other.
\begin{figure}[htbp]
    \centering
    \includegraphics[height=6.7cm]{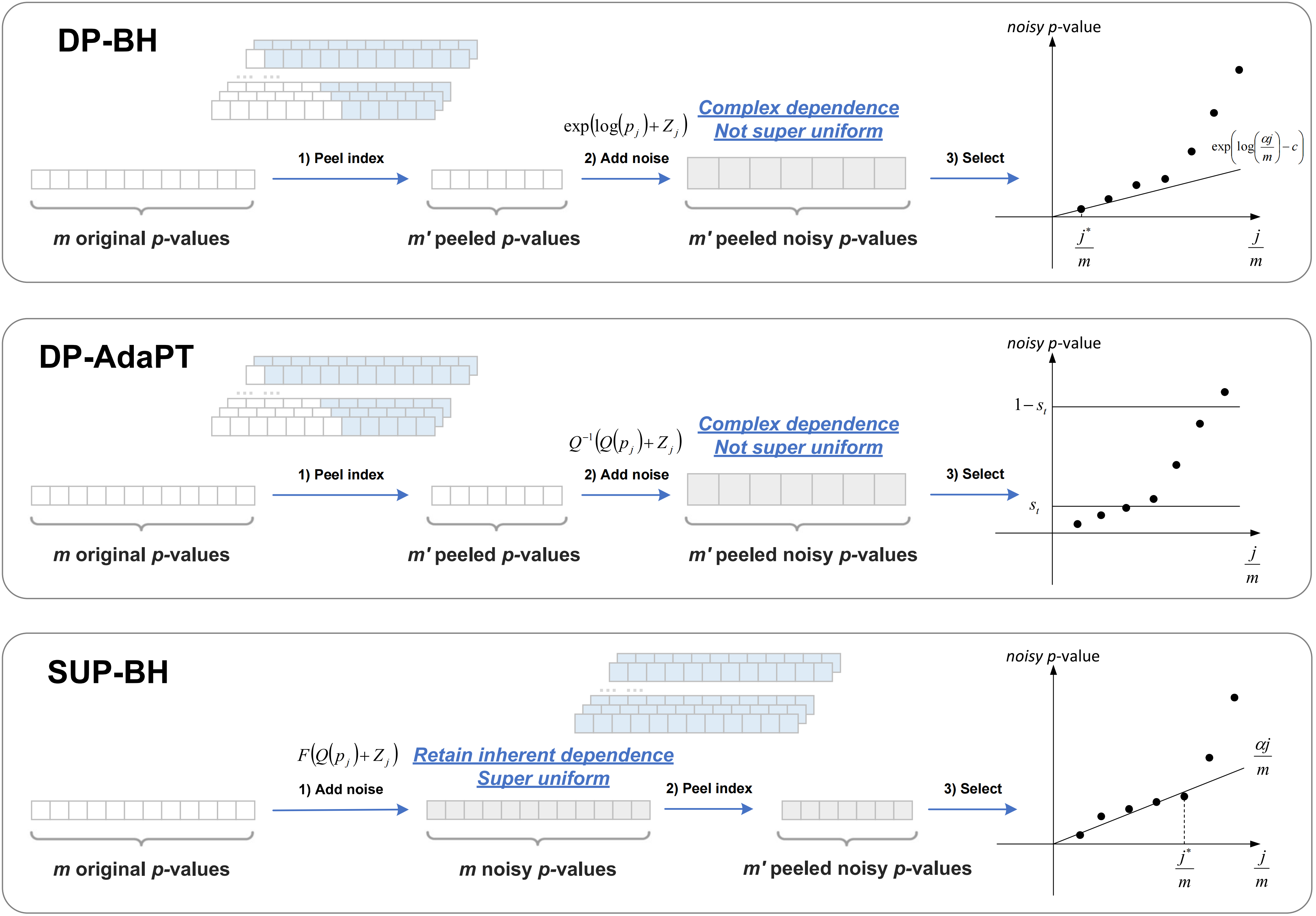}
    \caption{The flowcharts of DP-BH, DP-AdaPT, and our proposed SUP-BH.}
    \label{Fig: flowchart}
\end{figure}

Despite the maturity of multiple testing research, developing privacy-preserving multiple testing algorithms remains a non-trivial task \citep{dwork2021differentially,xia2023adaptive}, as large-scale $p$-values amplify the risk of disclosing sensitive information.
\citet{dwork2021differentially} developed differentially private Benjamini-Hochberg (DP-BH) and differentially private Bonferroni (DP-Bonf) procedures, assuming that the null $p$-values follow the uniform distribution $U(0,1)$. 
As illustrated in Figure~\ref{Fig: flowchart}, the DP-BH method consists of three steps. 
First, a randomized peeling algorithm is employed to reduce privacy budgets by peeling a small subset of numerous original $p$-values. Then, noise is added to the logarithms of these peeled $p$-values to ensure privacy. However, the first step gives rise to a post-selection obstacle: it introduces intricate dependence among the peeled $p$-values and destroys their super uniformity. Moreover, the random perturbation in the second step can lead to an unwarranted rejection of null hypothesis with large original
$p$-value. To establish theoretical results, \citet{dwork2021differentially} required that the largest original $p$-value among the rejected hypotheses can be bounded by a function of the number of rejections. Consequently, a conservative BH‐type threshold dependent on the privacy parameters is applied to the noisy statistic in the final step. This poses two critical limitations for the DP-BH method. First, it fails to overcome the theoretical challenge of controlling the commonly used FDR as defined in~\eqref{eq:Type-I error}, addressing only a generalized version of FDR. Second, its FDR control is conservative and sensitive to privacy parameters, which can reduce power, especially under stringent privacy requirements. The DP-Bonf is more conservative than DP-BH, exacerbating the second issue in terms of FWER control. 

On the other hand, \citet{xia2023adaptive} proposed the DP-AdaPT method, employing the mirror conservative condition rather than the super uniform condition. The DP-AdaPT method also comprises three steps, as depicted in Figure~\ref{Fig: flowchart}. First, it similarly utilizes a mirror peeling algorithm to reduce privacy costs. In the second step, noise is added to a function of the peeled $p$-values that preserves both privacy and the mirror conservative property. In the final step, symmetry and martingale techniques are utilized to achieve non-asymptotic FDR control.
While DP-AdaPT offers distinct advantages, it also inherits several limitations from the non-private AdaPT \citep{lei2018adapt}. First, it may struggle with FDR control when the null $p$-values are dependent. Second, AdaPT exhibits high variance in the false discovery proportion \citep{korthauer2019practical}, and this issue can be further exacerbated in DP-AdaPT due to the additional noise required for privacy. 
In addition, both DP-BH and DP-AdaPT face the limitation of requiring a pre-specified peeling number. Setting the peeling number too low relative to the number of signals substantially limits the power of multiple testing, whereas choosing a large peeling number amplifies the noise scale, thereby reducing power as well. The performance of both methods is highly sensitive to the peeling number, yet no established criterion exists for its determination. As highlighted by \citet{xia2023adaptive}, it remains an open problem to develop a differentially private procedure with finite-sample FDR control under the super uniform condition.

To address the challenges in private multiple testing, we propose a general private multiple testing framework under the super uniform condition, abbreviated as SUP. It is compatible with a range of privacy regimes, Type-I error metrics, and dependence structures. Departing from the forward design used in DP-BH and DP-AdaPT, the reversed design proposed in SUP guarantees the inferability for both Type-I error control and power analysis. Furthermore, we develop data-driven approaches for selecting the peeling number and boosting power. 
Specifically, this paper makes the following key contributions:
\vspace{-4pt}
\begin{enumerate}
    \item SUP includes a novel transformation of \( p \)-values, a reversed peeling algorithm, and a range of rejection thresholds.  By accounting for the noise distribution, this transformation ensures privacy while maintaining the super uniform property. The reversed peeling algorithm generates noisy \( p \)-values for all original \( p \)-values before peeling indices. By treating peeling and subsequent thresholding as a unified process, we circumvent the dependency issue among the peeled noisy $p$-values and achieve theoretical control of Type-I errors. Moreover, the reversed design enables us to establish an upper bound for the power of SUP and further quantify the privacy-induced power loss relative to the non-private benchmark. 
    
    \item SUP is the first privacy-preserving framework that allows practitioners to flexibly select rejection thresholds tailored to the desired Type-I error metric and the dependence structure of $p$-values. It supports various rejection thresholds, including SUP BH, SUP Benjamini-Yekutieli, SUP Bonferroni, and SUP Holm, and can easily incorporate others that rely on the super uniformity. Notably, these thresholds are privacy-parameter-free, providing stable and finite-sample Type-I error control across diverse privacy notions and arbitrary privacy levels. 
    
    \item SUP gains additional enhancements through adaptive estimators designed to capture signal information. These estimators assist in determining the peeling number and refining rejection thresholds, thereby significantly boosting power performance. Unlike estimators used in non-private settings, these estimators are specifically designed with low sensitivity to avoid excessive privacy costs. Furthermore, we theoretically demonstrate SUP's superiority over existing methods in terms of power, validated by extensive empirical analysis. 
\end{enumerate}

\vspace{-4pt}
The remainder of the paper is structured as follows. Section \ref{sec:meth} introduces the fundamental concepts of differential privacy and presents the SUP framework. Section \ref{sec:exten} discusses adaptive techniques for determining the peeling number and adjusting rejection thresholds to improve power. Section \ref{sec:theo} establishes both non-asymptotic and asymptotic theories for FDR and FWER control under the SUP framework, while comparing its power performance to its non-private counterpart and existing private methods. Extensive simulations and real data analysis in Sections \ref{sec:simu}-\ref{sec:real} demonstrate the numerical advantages of the SUP framework. Finally, Section \ref{sec:dis} concludes with a discussion on key questions for future research.

\section{Methodology}\label{sec:meth}

\subsection{Preliminaries}

In this section, we introduce some fundamental concepts of differential privacy. Consider a dataset \( D = \{x_1, \ldots, x_n\} \) comprising \( n \) samples from the universe \( \mathcal{X} \). A neighboring dataset $D^{\prime}=\left\{x_1^{\prime}, \ldots, x_n^{\prime}\right\}$ is identical to $D$ except for one sample. When the dataset $D$ contains sensitive information, directly releasing concerned statistics could compromise individual privacy. To mitigate risk, researchers employ a randomized mechanism $M$ that limits the difference between the distributions of $M(D)$ and $M\left(D^{\prime}\right)$, making it hard to infer whether any single individual's data is included in $D$ based on the mechanism's output. Denote $\mathcal{M}$ as the collection of all possible outputs of the mechanism $M$. \citet{dwork2014algorithmic} defines $(\varepsilon, \delta)$-differential privacy, abbreviated as $(\varepsilon, \delta)$-DP, to quantify this difference. 
\begin{definition}[Differential privacy]
    \label{def:dp}
    A randomized mechanism $M(\cdot)$ is $(\varepsilon, \delta)$-DP for some nonnegative $\varepsilon$ and $\delta$, if for all neighboring datasets $D$ and $D^{\prime}$, and for any measurable event $E \subset \mathcal{M}$, it follows that $
    \mathbb{P}(M(D) \in E) \leq e^\varepsilon \mathbb{P}\left(M\left(D^{\prime}\right) \in E\right)+\delta.$    
\end{definition}
In Definition~\ref{def:dp}, datasets $D$ and $D^{\prime}$ are considered fixed. 
The parameters \(\varepsilon\) and \(\delta\) determinate the level of privacy. Given a small value of \(\varepsilon\), \((\varepsilon, 0)\)-DP ensures that any output is almost equally likely across all neighboring databases. \((\varepsilon, \delta)\)-DP further relaxes the probabilistic difference between neighboring datasets. While ensuring \((\varepsilon, \delta)\)-DP makes it difficult for an attacker to ascertain whether a particular individual is included in dataset \( D \) based on the randomized output, its interpretability and versatility are limited.

Actually, from the perspective of an attacker, defining differential privacy through hypothesis testing is both natural and interpretable \citep{Kairouz2017the}. Specifically, consider the following hypothesis testing problem
\begin{equation}
    \label{eq:dp-to-test}
    H_0\text{: the underlying dataset is } D \;\text{ versus }\; H_1 \text{: the underlying dataset is } D^{\prime}\text {. }
\end{equation}
Denote \( x_i \) as an individual present solely in dataset \( D \) and not in \( D' \). Accepting the null hypothesis suggests that \( x_i \) is indeed present in dataset \( D \), whereas rejecting it implies that \( x_i \) is absent. From this perspective, \citet{dong2022gaussian} introduces $\mu$-Gaussian differential privacy, abbreviated as $\mu$-GDP, to strike a balance between interpretability and versatility.
\begin{definition}[Gaussian differential privacy]
    A mechanism $M$ is $\mu$-GDP if, for any given significance level $\alpha_0$, the power function of test~\eqref{eq:dp-to-test} does not exceed $1-G_\mu(\alpha_0)$, where $G_\mu(\alpha_0)=\Phi\left(\Phi^{-1}(1-\alpha_0)-\mu\right)$ and $\Phi(\cdot)$ is the cumulative distribution function of $N(0,1)$.
\end{definition}
$\mu$-GDP is parameterized by a single value $\mu$, which quantifies the difficulty of the hypothesis testing problem encountered by an attacker. Specifically, it measures the challenge of distinguishing between distributions $N(0,1)$ and $N(\mu, 1)$ using a single observation \citep{dong2022gaussian}. This straightforward characterization enhances the clarity and interpretability of privacy guarantee. Additionally, the definition of \(\mu\)-GDP ensures robust privacy protection even under multiple compositions of private mechanisms, making it particularly suitable for statistical methods that involve multiple or iterative operations. The composition theorem states that if two algorithms respectively provide $\mu_1$-GDP and $\mu_2$-GDP, their composition is $\sqrt{\mu_1^2+\mu_2^2}$-GDP. Additionally, the post-processing property of GDP asserts that for any deterministic function $f(\cdot)$ and $\mu$-GDP algorithm $M$, the post-processed algorithm $f \circ M$ is still $\mu$-GDP. As noted by \citet{dong2022gaussian}, $\mu$-GDP encompasses an infinite union of \((\varepsilon, \delta)\)-DP. Given these advantageous properties, we utilize $\mu$-GDP in the proposed framework.

A common approach to preserving privacy involves adding noise to the statistic of interest, with the magnitude of the noise calibrated to the sensitivity. The definition of sensitivity is as follows.
\begin{definition}[Sensitivity]
    \label{def:sen}
    Let $f: \mathcal{X}^n \rightarrow \mathbb{R}$ be a deterministic function from the dataset $D$ to $\mathbb{R}$. The local and global sensitivities of $f(\cdot)$ are respectively defined as
$$
\operatorname{LS}_{f(D)}=\sup _{\forall D^{\prime}}\left\|f(D)-f\left(D^{\prime}\right)\right\|, \quad \operatorname{GS}_f=\sup _{\forall D} \operatorname{LS}_{f(D)},
$$
where $D^{\prime}$ denotes a neighboring dataset of $D$, and $\|\cdot\|$ is the Euclidean norm.    
\end{definition}
The local sensitivity is determined by a specific dataset $D$, while the global sensitivity captures the maximum possible change resulting from modifying a single entry in $D$. To achieve $\mu$-GDP, we can apply the Gaussian mechanism to $f(D)$, with the noise scale calibrated to the global sensitivity $\operatorname{GS}_f$, as stated in Lemma~\ref{lemma:GDP}.
\begin{lemma}[Gaussian mechanism \citep{dong2022gaussian}]
    \label{lemma:GDP}
    The Gaussian mechanism that outputs $M(D) = f(D)+Z$ preserves $\mu$-GDP, where $Z$ follows $N\left(0, \operatorname{GS}_f^2 / \mu^2\right)$.
\end{lemma}


\subsection{Private multiple testing framework}

In this section, we present the SUP multiple testing framework, which simultaneously tests \( m \) hypotheses while controlling Type-I errors at a significance level \(\alpha\) and ensuring privacy. Our framework operates on a set of $p$-values $\mathcal{P} = \{p_j\}_{j=1}^m$ and comprises three key steps: generating noisy $p$-values, peeling a subset of noisy $p$-values, and post-selecting hypotheses based on various thresholds. 

In large-scale multiple testing, the $p$-value, denoted as $p(D)$, is a commonly concerned statistic, where $p(\cdot)$ is considered a deterministic real-valued function of the dataset. For a given $p$-value function, the local sensitivity $\operatorname{LS}_{p(D)}$ varies significantly over the range $p(D) \in [0,1]$. Consequently, adding noise directly to the \(p\)-value with a noise scale calibrated to the global sensitivity could result in excessive privacy protection, potentially overwhelming the signals. To address this challenge, \citet{dwork2021differentially} and \citet{xia2023adaptive} propose adding noise to its logarithm and the test statistic used to obtain the $p$-value, respectively. 
In this paper, we consider adding noise to a general transformation of $p$-value, $Q(p(D))$, abbreviated as $Q_p(D)$.
The function $Q_p(D)$ should meet the following criteria. First, \(\operatorname{GS}_{Q_p}\) should tend to 0 as the sample size approaches infinite and be computationally efficient to facilitate practical implementation. Second, the local sensitivity \(\operatorname{LS}_{Q_p(D)}\) should closely approximate the global sensitivity \(\operatorname{GS}_{Q_p}\) across all values of $Q_p(D)$. A quantile function is particularly well-suited for these purposes, with examples of calculating \(\operatorname{GS}_{Q_p}\) provided in \citet{xia2023adaptive}.
Given a specific function $Q$, the noisy $p$-value $\tilde{p}$ can be generated as follows:
\begin{equation}
    \label{eq:noisy_p}
    \tilde{p} = F\Big(Q\big(p(D)\big) + Z\Big),
\end{equation}
where $Z$ is independently sampled from a normal distribution $N(0,\sigma^2)$ to guarantee privacy and $F$ represents the cumulative distribution function of $Q(p_{U}) + Z$, with $p_{U}$ following the uniform distribution $U(0,1)$. While we introduce Gaussian noise to satisfy GDP, it can also be replaced with Laplace noise to ensure $(\varepsilon,\delta)$-DP, as detailed in Appendix B. The standard deviation $\sigma$ depends on the privacy parameters and the global sensitivity $\operatorname{GS}_{Q_p}$.
The function $F$ is specifically designed to correct the variance introduced by $Z$, ensuring that the transformed noisy $p$-value $\tilde{p}$ retains the super uniform property, as detailed in Lemma~\ref{lemma:super_u}. 
\begin{lemma}
    \label{lemma:super_u}
    For any $p$-value $p$ that satisfies the super uniform property, denoted as $\mathbb{P}(p \leq t) \leq t$ for any $ t \in [0,1]$, the transformed noisy $p$-value $\tilde{p}$, obtained via equation~\eqref{eq:noisy_p}, still satisfies the super uniform property.
\end{lemma}
The first step of SUP generates \( 1 + m' \) sets of noisy \( p \)-values, where \( m' \) is a pre-specified peeling number. These sets are denoted as \( \tilde{\mathcal{P}}^{(k)} = \{\tilde{p}^{(k)}_j\}_{j=1}^{m} \), with \( \tilde{p}_j^{(k)} = F\left(Q(p_j) + Z_j^{(k)}\right) \) for \( k = 0, 1, \dots, m' \). We set the standard deviation of \( Z_j^{(0)} \) to \( \sigma_0 \), and for \( k \geq 1 \), the standard deviation of \( Z_j^{(k)} \) to \( \sigma_1 \). The set \( \tilde{\mathcal{P}}^{(0)} \) is used to control Type-I errors in the final step, while the remaining sets are employed for peeling index. Since each noise term \( Z_j^{(k)} \) is generated independently, the dependence among the noisy \( p \)-values is weaker than that of the original \( p \)-values. Theoretical analysis based on all noisy \( p \)-values effectively addresses the dependency issue among the peeled noisy \( p \)-values encountered by \citet{dwork2021differentially}.

We proceed to discuss the rationale and methodology for peeling a subset of noisy $p$-values with much smaller cardinality, denoted as $m' \ll m$. If we ensure the privacy of all $p$-values, the composition theorem \citep{dong2022gaussian} indicates that the standard deviation of the noise should be proportional to the square root of the total number of reported hypotheses. When the number of hypotheses is large, reporting all $p$-values with a privacy guarantee would significantly amplify the noise associated with each $p$-value, thereby obscuring signals and reducing the efficacy of multiple testing. Thus, it is more feasible to peel a subset of $p$-values, prioritizing those most likely to be signals.

\begin{algorithm}
    \setstretch{0.9}
    \caption{The Reversed Peeling Algorithm}
    \label{alg:peeling}
    \begin{algorithmic}[1]
    \Require $p$-values $\{p_j\}_{j=1}^m$ with global sensitivity at most $\operatorname{GS}_{Q_p}$, peeling number $m'$, noise scales $\sigma_0$ and $\sigma_1$.
    \State generate noisy $p$-value sets $\tilde{\mathcal{P}}^{(k)}$ for $k = 0, 1, \dots, m'$;
    \State set initial index set $S = \{1, \dots, m\}$;
    \For{$k = 1$ to $m'$}
        \State let $j_k = \arg\min_{j \in S} \tilde{p}^{(k)}_j$;
        \State update $S = S \setminus \{j_k\}$;
    \EndFor
    \Ensure the peeled indices $\mathcal{J} = \{j_1, \dots, j_{m'}\}$ and $p$-values $\{\tilde{p}^{(0)}_j: j \in \mathcal{J}\}$.
    \end{algorithmic}
\end{algorithm}

In the second step, we peel the index \(j_k\) corresponding to the minimum noisy \(p\)-value among \(\tilde{\mathcal{P}}^{(k)}\) for each \(k\) from 1 to \(m'\). Once an index is peeled, it can not be peeled again. The peeled indices $\mathcal{J} = \{j_1, \dots, j_{m'}\}$ and the corresponding $p$-values $\{\tilde{p}_j^{(0)}: j \in \mathcal{J}\}$ are subsequently utilized in the third step. The first two steps are encapsulated in  Algorithm~\ref{alg:peeling}. Guidance on choosing the peeling number $m'$ is provided in Section~\ref{sec:exten}. The privacy of Algorithm~\ref{alg:peeling} is established in Lemma~\ref{lemma:peeling-DP}, which follows from Lemma~\ref{lemma:GDP} and the composition theorem \citep{dong2022gaussian}.

\begin{lemma}[Privacy guarantee]
    \label{lemma:peeling-DP}
        The Reversed Peeling Algorithm~\ref{alg:peeling} achieves $\mu$-GDP with
        $\sigma_0 = \sqrt{2m'}\operatorname{GS}_{Q_p}/\mu$ and $\sigma_1 = 2\sqrt{2m'}\operatorname{GS}_{Q_p}/\mu$.
\end{lemma}
\begin{remark}[Reversed design]
    Although both our reversed peeling Algorithm \ref{alg:peeling} and the forward peeling Algorithm \ref{alg:Peeling_Mechanism} of \citet{dwork2021differentially} use peeling to economize on the privacy budget, the forward design yields peeled noisy $p$-values exhibiting complex dependence and lacking super uniformity. This forces one to adopt conservative rejection thresholds for Type-I error control. By contrast, our reversed design employs a general transformation to generate all $m$ noisy $p$-values before peeling index, thereby retain inherent dependence of original $p$-values and the super uniformity. Basing the analysis on all $m$ noisy $p$-values, rather than on the peeled subset, allows us to construct privacy-parameter-free rejection thresholds and thereby achieve robust Type-I error control. Moreover, as demonstrated in Section~\ref{sec:theo}, the reversed design enables us to quantify the power loss attributable to privacy constraints. 
\end{remark}

The third step of our framework focuses on post-selecting a subset of peeled noisy \( p \)-values. This selection is based on tailored thresholds designed to control different Type-I errors. To enhance clarity, we first present the SUP Multiple Testing Algorithm~\ref{alg:SUP} and provide details on specific thresholds in the next section.
\begin{algorithm}
    \setstretch{0.9}
    \caption{The SUP Multiple Testing Algorithm}
    \label{alg:SUP}
    \begin{algorithmic}[1]
    \Require target FDR level $\alpha$, $p$-values $\{p_j\}_{j=1}^m$ with global sensitivity at most $\operatorname{GS}_{Q_p}$, privacy parameter $\mu$, peeling number $m'$, noise scales $\sigma_0$ and $\sigma_1$, thresholds $\lambda_1, \dots, \lambda_{m'}$, step-up parameter $\zeta \in \{0,1\}$.
    \State apply Algorithm~\ref{alg:peeling} to obtain $\mathcal{J} = \{j_1, \dots, j_{m'}\}$ with $\sigma_0 = \sqrt{2m'}  \operatorname{GS}_{Q_p} / \mu$ and $\sigma_1 = 2\sqrt{2m'} \operatorname{GS}_{Q_p} / \mu$;
    \State sort noisy $p$-values $\{\tilde{p}_j^{(0)}: j \in \mathcal{J}\}$ in increasing order: $\tilde{p}_{(1)}^{\mathcal{J}} \leq \tilde{p}_{(2)}^{\mathcal{J}} \leq \dots \leq \tilde{p}_{(m')}^{\mathcal{J}}$;
    \If{$\zeta = 1$}
    \State $j^{\star} = \sup\{j: \tilde{p}_{(j)}^{\mathcal{J}} \leq \lambda_{j}, j \leq m'\}$;
    \Else
    \State $j^{\star} = \inf\{j: \tilde{p}_{(j)}^{\mathcal{J}} > \lambda_{j}, j \leq m'\} - 1$;
    \EndIf
    \Ensure the selected indices $\mathcal{J}^{\star} = \{(1), \dots, (j^{\star})\}$ and selected $p$-values $\{\tilde{p}_j^{(0)}: j \in \mathcal{J}^{\star}\}$; if $j^\star \not\in \mathbb{R}^+$, output $\varnothing$.
    \end{algorithmic}
\end{algorithm}

\begin{theorem}[Privacy guarantee of SUP]
    \label{Theo:SUP-DP}
    The Algorithm~\ref{alg:SUP} is $\mu$-GDP. 
\end{theorem} 
Since the third step can be regarded a deterministic function of the outcomes from the first two steps, it does not reveal any additional privacy. Consequently, Theorem~\ref{Theo:SUP-DP} follows directly from Lemma~\ref{lemma:peeling-DP}, combined with the post-processing property of GDP. 
It is worth noting that, based on the relation between $\mu$-GDP and $(\varepsilon, \delta)$-DP, Algorithm~\ref{alg:SUP} achieves $(\varepsilon, \delta)$-DP for all $\varepsilon > 0$ where $\delta = \Phi(-\varepsilon / \mu + \mu / 2) - e^\varepsilon \Phi(-\varepsilon / \mu - \mu / 2)$. Furthermore, Algorithm~\ref{alg:SUP} can be extended to maintain a specific $(\varepsilon, \delta)$-DP by replacing the Gaussian noise with the Laplace noise, as detailed in Supplementary Material~\ref{sup:laplace}.
\vspace{-1ex}
\subsection{Controlling Type-I errors with tailored thresholds}\label{sec:thresholds}
Most existing methods for controlling Type-I errors with the super uniformity begin by ordering the $p$-values and then applying either a step-up or step-down procedure, with threshold tailored to the target error metric. A step-up procedure progresses from larger to smaller $p$-values, while a step-down procedure proceeds in the opposite direction. Stopping criteria differ between the two approaches.

In what follows, we design thresholds for SUP to control the FDR and the FWER. Specifically, we sort the peeled noisy $p$-values $\{\tilde{p}_j^{(0)}: j \in \mathcal{J} \}$ in increasing order, yielding $\tilde{p}_{(1)}^{\mathcal{J}} \leq \tilde{p}_{(2)}^{\mathcal{J}} \leq \dots \leq \tilde{p}_{(m')}^{\mathcal{J}}$. For each $\tilde{p}_{(j)}^{\mathcal{J}}$, we introduce a corresponding threshold $\lambda_{j}$, for $j = 1, \dots, m'$. In terms of FDR control, we employ two tailored thresholds: $\lambda_j = \alpha j / m$ (SUP BH) and $\lambda_j = \alpha j / (\sum_{l=1}^m {m}/{l})$ (SUP Benjamini-Yekutieli).
The SUP BH threshold is suitable when the $p$-values are independent or exhibit certain forms of positive dependence. By contrast, the SUP Benjamini-Yekutieli threshold is more conservative and thus appropriate for scenarios involving arbitrary dependence among the $p$-values. Both thresholds employ a step-up procedure, sequentially selecting \(p\)-values up to \( \tilde{p}_{(j^{\star})}^{\mathcal{J}} \), where \( j^{\star} = \sup\{j: \tilde{p}_{(j)}^{\mathcal{J}} \leq \lambda_{j}, j \leq m'\} \).
For FWER control, we also introduce two thresholds: $\lambda_j = \alpha / m$ (SUP Bonferroni) and $\lambda_j = \alpha / (m + 1 - j)$ (SUP Holm).
Both the SUP Bonferroni and SUP Holm thresholds are designed to handle arbitrary dependent \(p\)-values. The SUP Bonferroni threshold remains constant and yields identical results whether a step-up or step-down procedure is employed. In contrast, the SUP Holm threshold is better suited to a step-down procedure, where \(p\)-values are selected from \( \tilde{p}_{(1)}^{\mathcal{J}} \) to \( \tilde{p}_{(j^{\star})}^{\mathcal{J}} \), with \( j^{\star} = \inf\{j: \tilde{p}_{(j)}^{\mathcal{J}} > \lambda_{j}, j \leq m'\} - 1 \).
These thresholds exemplify only a portion of the methods available for controlling Type-I errors. There are many methods that utilize the super uniform property to control Type-I errors, such as Hochberg's method \citep{hochberg1988sharper}, can also be integrated into the SUP framework.

\vspace{-2ex}
\section{Adaptive SUP multiple testing}\label{sec:exten}

In this section, we explore adaptive techniques to capture structured information of signals, which aids in determining the peeling number and more powerful thresholds. 

\subsection{Adaptive peeling number}

We first discuss the impact of peeling number on the Type-I error and power. So far, all private multiple testing methods, including our SUP framework, the DP-BH method \citep{dwork2021differentially}, and the DP-AdaPT \citep{xia2023adaptive}, rely on a pre-specified peeling number $m'$. The choice of \(m'\) is pivotal in influencing the noise variance and the conservatism of rejection thresholds, especially in the DP-BH method. In experiments, \citet{dwork2021differentially} employs a relatively modest peeling number \(m' = 100\). While this modest peeling number contributes to a lower noise variance, it limits the maximum power to \(m'/m_1\) when the number of non-null hypotheses $m_1$ exceeds $m'$. To address this limitation, \citet{xia2023adaptive} recommends selecting a slightly larger \(m'\). However, excessively increasing \(m'\) amplifies the noise variance, complicating the distinction between signal and noise, resulting in a significant decrease in power. At present, no research offers effective criteria to select the peeling number in private multiple testing.

\begin{figure}[htbp]
    \centering
    \includegraphics[height=6.5cm]{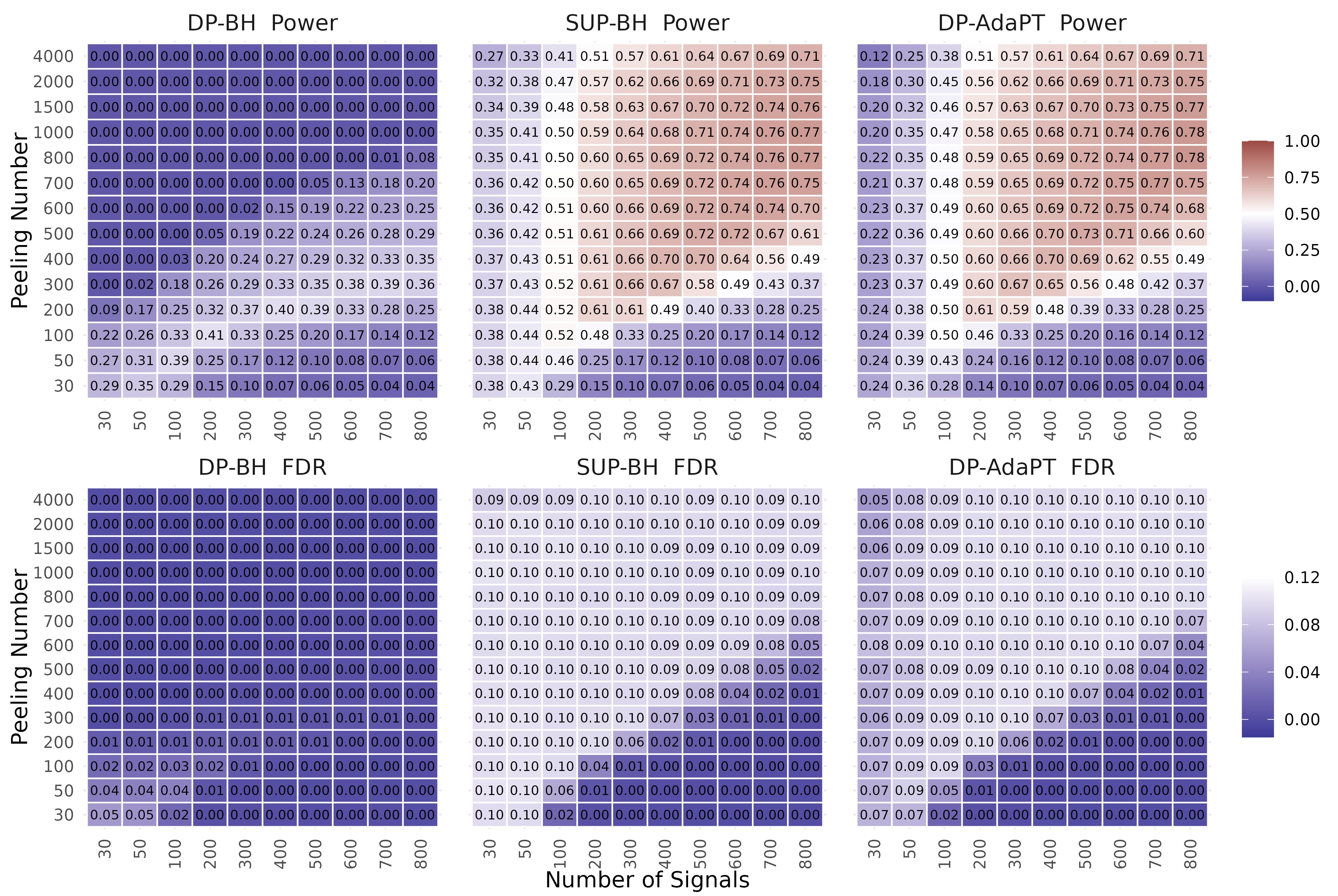}
    \caption{\textit{The FDRs and powers for DP-BH, DP-AdaPT, and Algorithm~\ref{alg:SUP} with the SUP BH threshold across varying numbers of signals, peeling numbers, and medium privacy requirement.}}
    \label{Fig: Heatmap_mid}
\end{figure}

To analyze the impact of the peeling number on Type-I error and power, we conducted simulations by varying the peeling number, the number of true signals, and the privacy parameter $\varepsilon$. Detailed simulation settings are provided in Supplementary Material~\ref{sup:numer}. Figures~\ref{Fig: Heatmap_mid}, ~\ref{Fig: Heatmap_low}, and \ref{Fig: Heatmap_high} display results under medium, low, and high noise variances, respectively. Our findings indicate that a small peeling number leads to overly conservative outcomes for all methods. Conversely, a large peeling number reduces power for all methods and results in conservative Type-I error control, particularly for the DP-BH method. Satisfactory performance across all methods is achieved when the peeling number matches or slightly exceeds the number of signals \(m_1\). The influence of the peeling number on power becomes more pronounced as the noise variance increases. Based on these observations, we aim to choose the peeling number adaptively by estimating \(m_1\).

Denote the number and proportion of null hypotheses as $m_0$ and $\pi_0 = m_0/m$, respectively. \citet{storey2002direct} proposed a widely adopted estimator of $\pi_0$, given by:
\begin{equation}
    \label{eq:pi_0_storey}
    \tilde{\pi}_0=\frac{\sum_{j=1}^m \mathbbm{1}(p_j>\tau)}{m(1-\tau)}.    
\end{equation}
Intuitively, if each null $p$-value $p_j$ follows the uniform distribution $U(0,1)$, there should be approximately $\pi_0 m(1-\tau)$ $p$-values greater than $\tau$. Here, \(\tau\) is a predefined cutoff greater than \(\alpha\), typically set to \(\tau = 0.5\) or \(\tau = 0.75\), ensuring that most \( p \)-values above \(\tau\) are likely to correspond to null hypotheses. Then, \(m(1-\tilde{\pi}_0)\) can serve as an estimate for \(m_1\), to which Gaussian noise is added for privacy protection. However, the global sensitivity of \(\tilde{\pi}_0\) is \(\operatorname{GS}_{\tilde{\pi}_0} = (1-\tau)^{-1}\), which does not depend on the sample size $n$ and is excessively large for \(\tilde{\pi}_0\). Adding noise calibrated to $\operatorname{GS}_{\tilde{\pi}_0}$ leads to a highly unstable estimate of $\pi_0$. 
This high sensitivity arises from the discontinuity of the indicator function $\mathbbm{1}(p_j>\tau)$.
To address this issue, a continuous function $\mathbbm{1}(p_j>\tau)\left(Q(p_j) - Q(\tau)\right)$ is employed to design a novel estimate of $\pi_0$. Beyond reducing sensitivity, consistency of the estimator is crucial. Let $S_0$ and $S_1$ be the index set of null and non-null hypotheses, respectively. Notice that if $p_j$ follows the uniform distribution $U(0,1)$ for $j \in S_0$ and tends to be small for $j \in S_1$, the sum $\sum_{j=1}^m \mathbbm{1}(p_j>\tau)\left(Q(p_j) - Q(\tau)\right)$ approximates $\sum_{j\in S_0} \mathbbm{1}(p_j>\tau)\left(Q(p_j) - Q(\tau)\right)$, serving as an estimate for $\pi_0 m(1-\tau) E_{\tau}$. Here, $E_{\tau} = \mathbb{E}\left[Q(p_U) - Q(\tau) \mid p_U > \tau\right]$, with $p_U$ following $U(0,1)$. Thus, we propose to estimate $\pi_0$ by
\begin{equation}
    \label{eq:pi_0_dagger}
    \bar{\pi}_0 = \frac{\sum_{j=1}^m \mathbbm{1}(p_j>\tau)\left(Q(p_j) - Q(\tau)\right)}{m(1-\tau)E_{\tau}}.
\end{equation}
We derive that $\bar{\pi}_0$ has a low global sensitivity $\operatorname{GS}_{\bar{\pi}_0 } = \operatorname{GS}_{Q_p}(1-\tau)^{-1}E_{\tau}^{-1}$, which approaches 0 as the sample size tends to infinite. 
To ensure privacy, a noisy estimate of $\pi_0$ can be obtained by adding a Gaussian noise $Z_{\bar{\pi}_0}$ to $\bar{\pi}_0$. It is clear that the noise introduced here is considerably smaller than the noise added in Algorithm~\ref{alg:SUP}, whose scale grows proportionally with the peeling number by the composition theorem. Therefore, as long as the privacy constraint is not excessively strict, this perturbation minimally affects the estimation accuracy.
Subsequently, the peeling number can be determined as
\begin{equation}
    \label{eq:m_dagger}
    m^{\dagger} = \max\left\{ \lceil(1+c)m(1-\bar{\pi}_0  + Z_{\bar{\pi}_0})\rceil, \tilde{m}\right\},
\end{equation}
where $\lceil\cdot\rceil$ represents rounding up to the nearest integer, and $\tilde{m}>0$ and $c\geq0$ are user-specified parameters providing flexibility. Specifically, choosing $\tilde{m}$ as a small constant ensures effectiveness in cases where $\bar{\pi}_0$ is approximately 1, thereby preventing $m^{\dagger}$ from becoming too small. We recommend selecting $c = (1-\alpha)^{-1} -1$, since if we aim to identify all signals while exactly controlling the false discovery proportion at level $\alpha$, the minimal required peeling number can be derived as $(1-\alpha)^{-1}m(1-\pi_0)$.
Then, the data-driven peeling number \( m^{\dagger} \) can replace the fixed \( m' \), and each SUP threshold can be adjusted accordingly to \(\lambda_j\), \(j = 1, \dots, m^{\dagger}\).

\subsection{Adaptive thresholds}

Although we have derived a data-driven peeling number \( m^{\dagger} \), there remains an opportunity to further enhance the power of SUP. As demonstrated in Theorems~\ref{Theo:SUP-FDR}-\ref{Theo:FWER}, Algorithm~\ref{alg:SUP}, employing the SUP BH and SUP Bonferroni thresholds, only conservatively maintain Type-I errors at \(\pi_0 \alpha\). Multiplying these thresholds by \(\pi_0^{-1} \) enables exact Type-I error control at \(\alpha\) and significantly boosts power in dense signal scenarios. 
In contrast, Algorithm~\ref{alg:SUP} with the SUP Holm threshold precisely controls Type-I error at the target level. It can only be improved by adapting the data-driven peeling number \( m^{\dagger} \), without requiring any scale correction to the threshold.

In non-private settings, some methods \citep{storey2002direct, li2019multiple, ignatiadis2021covariate} modify rejection thresholds by incorporating structured signal information.
For example, \citet{storey2002direct} suggests using \(\tilde{\pi}_0^{-1} \alpha j / m\) instead of the standard BH threshold \(\alpha j / m\). Nevertheless, the proposed private estimator $\bar{\pi}_0 + Z_{\bar{\pi}_0}$, where $Z_{\bar{\pi}_0}$ is the added noise to ensure privacy, is solely suitable for determining an adaptive peeling number and cannot be used to adjust the threshold. This limitation arises because the theoretical guarantee of Type-I error control requires the expectation of the estimator of $\pi_0^{-1}$ closely approximates $\pi_0^{-1}$, which is not achievable for $\mathbb{E}\left[1/\left(\bar{\pi}_0 + Z_{\bar{\pi}_0}\right)\right]$. 

To achieve effective Type-I error control with an adaptive threshold, it is preferable to add noise $Z_{\tau}$ to $\bar{\pi}_0^{-1}$ rather than directly to $\bar{\pi}_0$, ensuring that $\mathbb{E}\left[\bar{\pi}_0^{-1} + Z_{\tau}\right] = \mathbb{E}\left[\bar{\pi}_0^{-1}\right] + 0 \approx \pi_0^{-1}$. Note that the denominator $\sum_{j=1}^m \mathbbm{1}(p_j>\tau)\left(Q(p_j) - Q(\tau)\right)$ of $\bar{\pi}_0^{-1}$ can be 0. To prevent the global sensitivity of $\bar{\pi}_0^{-1}$ from becoming infinity, we propose a refined estimator 
\begin{equation}
    \label{eq:pi0_dot}
    \bar{\pi}_0^{-1} = \frac{m(1-\tau)E_{\tau}}{\max\left\{\sum_{j=1}^m \mathbbm{1}(p_j>\tau)\left(Q(p_j) - Q(\tau)\right), c_0 m(1-\tau)E_{\tau}\right\}},
\end{equation}
where $c_0$ represents the minimum assumed proportion of null hypotheses, ensuring $\bar{\pi}_0 \geq c_0$. A typical choice is $c_0 = 0.5$. The global sensitivity of $\bar{\pi}_0^{-1}$ is derived as:
\begin{equation}
    \label{eq:sen_s_tau}
    \operatorname{GS}_{\bar{\pi}_0^{-1}} = c_0^{-1} - \left(c_0 + \frac{\operatorname{GS}_{Q_p}}{(1-\tau)E_{\tau}}\right)^{-1}. 
\end{equation}
Using this sensitivity, a private estimate of $\pi_0$ can be obtained as
\begin{equation}
    \label{eq:pi0_hat}
    \hat{\pi}_0 = \left(\bar{\pi}_0^{-1} + Z_{\tau}\right)^{-1},
\end{equation}
where $Z_{\tau}$ follows the normal distribution $N(0, \sigma_{\tau}^2)$. To achieve $\mu$-GDP for reporting $\hat{\pi}_0$, the standard deviation $\sigma_{\tau}$ can be determined as $\operatorname{GS}_{\bar{\pi}_0^{-1}}/\mu$.
In this case, we can adopt an adaptive threshold $\tilde{\lambda}_j = {\lambda_j}/{\hat{\pi}_0}.$
By leveraging the post-processing property of GDP, a corresponding adaptive peeling number \( m^{\star} \) can be used in conjunction with the adaptive threshold to avoid unnecessary privacy costs, which is defined as $m^{\star} = \max\left\{ \lceil(1+c)m(1-\hat{\pi}_0)\rceil, \tilde{m}\right\}.$
We summarize the procedure for jointly adapting the peeling number \( m^{\star} \) and threshold $\tilde{\lambda}_j$ in Algorithm~\ref{alg:AdaSUP} in Supplementary Material~\ref{sec:A}, which ensures more exact Type-I error control and boosts power. 
\begin{theorem}[Privacy guarantee of Adaptive SUP]
    \label{Theo:ASUP-DP}
    The Algorithm~\ref{alg:AdaSUP} is $\mu$-GDP.
\end{theorem} 

\vspace{-2ex}
\section{Theoretical results}\label{sec:theo}
In this section, we first develop non-asymptotic theories for controlling FDR and FWER across various thresholds within the SUP framework, and establish asymptotic theories for the Adaptive SUP framework. We then quantify the privacy-induced power loss of SUP relative to its non-private counterpart and demonstrate the power superiority of SUP over existing privacy-preserving methods.

\subsection{Type-I errors control}
The primary challenge in controlling Type-I errors arises from the privacy-preserving peeling algorithm \citep{dwork2006calibrating}. It necessitates post-selection inference based on peeled \( p \)-values, whose complex distribution and dependence structure render theoretical analysis highly intricate. In contrast, by leveraging carefully designed algorithms and artful theoretical techniques, we overcome this challenge effectively. Theorem~\ref{Theo:SUP-FDR} demonstrates that Algorithm~\ref{alg:SUP}, employing the SUP BH and SUP Benjamini-Yekutieli thresholds, non-asymptotically controls the FDR at the target level under independence and arbitrary dependence conditions, respectively. Meanwhile, Theorem~\ref{Theo:FWER} confirms that with the SUP Bonferroni and SUP Holm thresholds, Algorithm~\ref{alg:SUP} non-asymptotically controls the FWER at the target level under arbitrary dependency. It is worth noting that the SUP framework does not impose any additional conditions beyond those required by the corresponding non-private methods.
\begin{theorem}[FDR control]
    \label{Theo:SUP-FDR}
    Assume that the null $p$-values are super uniform.
    \begin{enumerate}[(1).]
        \item If the null $p$-values are independent of each other and of all non-null $p$-values, then Algorithm~\ref{alg:SUP}, employing the SUP BH threshold and the step-up parameter $\zeta = 1$, controls the FDR at a level less than or equal to $\pi_0\alpha$.
        \item Algorithm~\ref{alg:SUP}, employing the SUP Benjamini-Yekutieli threshold and the step-up parameter $\zeta = 1$, always controls the FDR at level less than or equal to $\pi_0\alpha$.
    \end{enumerate}
\end{theorem}
In the non-private setting, it is well known that the BH procedure controls the FDR not only under independence but also under positive regression dependence on a subset \citep[PRDS,][]{benjamini2001control, finner2009false}. 
Although the PRDS condition is not applicable in the private setting, Remark~\ref{remark:prds} in Supplementary Material~\ref{sup:proof} shows that Algorithm~\ref{alg:SUP} with BH threshold still controls FDR under an analogous positive dependence condition.

\begin{theorem}[FWER control]
    \label{Theo:FWER}
    Assume that the null $p$-values are super uniform.
    \begin{enumerate}[(1).]
        \item Algorithm~\ref{alg:SUP}, using the SUP Bonferroni threshold, always controls the FWER at a level less than or equal to $\pi_0\alpha $.
        \item Algorithm~\ref{alg:SUP}, employing the SUP Holm threshold and the step-up parameter $\zeta = 0$, always controls the FWER at a level less than or equal to $\alpha$.
    \end{enumerate}
\end{theorem}

We proceed to consider the theoretical property of the Adaptive SUP framework. 
Denote $V_{\tau} = \#\{j: p_j > \tau, \tilde{p}_j^{(0)} \text{ is rejected}, j \in S_0\}$, where $\tau$ is the same one emplyed in \eqref{eq:pi0_dot}. Let $\epsilon_{1,\tau} = \mathbb{E}[V_{\tau}/(R\vee 1)]$ and $\epsilon_{2,\tau} = \mathbb{P}(V_{\tau} >0)$.
The following theorem establishes the validity for FDR and FWER control achieved by Algorithm~\ref{alg:AdaSUP}. 
\begin{theorem}[Adaptive Type-I errors control]
    \label{Theo:Ada-SUP}
    Assume that all null $p$-values are super uniform, possess non-decreasing density functions, and are independent of each other as well as of all non-null $p$-values. Additionally, for $j \in S_0$, assume that $Q(p_j)$ has finite expectation and variance.
    \begin{enumerate}[(1).]
        \item Algorithm~\ref{alg:AdaSUP}, employing the Adaptive SUP BH threshold and the step-up parameter $\zeta = 1$, asymptotically controls the FDR at a level less than or equal to $\alpha + \epsilon_{1,\tau} $.
        \item Algorithm~\ref{alg:AdaSUP} with the Adaptive SUP Bonferroni threshold  asymptotically controls the FWER at a level less than or equal to $\alpha + \epsilon_{2,\tau} $.
    \end{enumerate}
\end{theorem}
In practical applications, the Type-I error level $\alpha$ is typically much smaller than \(\tau\); for instance, \(\alpha = 0.1\) and \(\tau = 0.5\). We remark here that the two terms $\epsilon_{1,\tau}$ and $\epsilon_{2,\tau}$ are challenging to eliminate in the private scenario. This difficulty arises because, in the non-private scenario, removing these terms relies on the idea of \(\tau\)-censoring \citep{li2019multiple, ignatiadis2021covariate}, meaning that the rejected \(p\)-values must be smaller than \(\tau\), which ensures that these \(p\)-values do not overlap with those used to calculate the weight, such as the estimator of $\pi_0^{-1}$, thereby guaranteeing Type-I error control.
In contrast, due to the added noise, it becomes theoretically possible for $\tilde{p}^{(0)}_j$ to exceed $\tau$ and still lead to the rejection of the $j$-th hypothesis. Nevertheless, these two terms are intuitively negligible compared to $\alpha$. This is because our algorithm tends to select indices corresponding to smaller original $p$-values, making it unlikely for indices associated with $p$-values greater than $\tau$ to be peeled when the scale of added noise is moderate. On the other hand, when the noise scale becomes large, the noisy $p$-values corresponding to the non-null and null hypotheses become indistinguishable, leading to very few or no rejections. 
This intuition is further supported by our simulation results. In addition, we note that the non-decreasing density assumption guarantees that $\mathbb{E}\left[\hat{\pi}_0^{-1}\right]$ does not exceed $\pi_0^{-1}$ asymptotically, ensuring reliable Type-I error control. It holds in general scenarios, including when null $p$-values follow $U(0,1)$.
The SUP Benjamini-Yekutieli threshold accommodates arbitrary dependency structures. However, ensuring the aforementioned inequality under such general conditions is challenging and warrants further investigation. Under the independence assumption in Theorem~\ref{Theo:Ada-SUP}, the SUP BH threshold is more powerful than the SUP Benjamini-Yekutieli threshold, eliminating the need for adjusting the latter.


\vspace{-6pt}
\subsection{Power analysis}
Thus far, we have established that Type-I errors can be controlled under privacy constraints. The next critical question concerns the impact of privacy on power. However, the peeling mechanisms employed in \citet{dwork2021differentially,xia2023adaptive} increase algorithmic complexity and induce intricate dependencies among the peeled noisy $p$-values, rendering a rigorous power analysis non-trivial. To the best of our knowledge, no existing study has yet quantified the privacy-induced power loss of multiple testing procedures relative to their non-private counterparts, given the same target Type-I error level.


Nonetheless, we observe that the proposed reversed peeling algorithm facilitates not only the analysis of Type-I error control but also a clear decomposition of privacy-induced power loss. For our proposed framework, the loss in power can be partitioned into two components. The first component is attributable to the noise introduced into the inference $p$-values $\tilde{p}^{(0)}_1,\dots,\tilde{p}^{(0)}_{m}$, which diminishes the distinction between the null and alternative distributions. This second component of the power reduction arises from the random peeling of indices based on the noisy sets $\tilde{\mathcal P}^{(k)}$ for $k = 1,\dots,m'$, which may discard true signals.
To connect our proposed framework with its non-private counterpart, we introduce an intermediate Algorithm \ref{alg:SUP-Truncated} in Supplementary Material~\ref{sec:A}, which serves as a bridge solely for theoretical analysis and does not offer desired privacy guarantees. Let $\tilde{p}^{(0)}_{(j)}$ be the $j$-th order statistic of noisy $p$-values $\tilde{p}^{(0)}_{1}, \dots, \tilde{p}^{(0)}_{m}$. Algorithm~\ref{alg:SUP-Truncated} omits the random peeling step and instead applies the rejection thresholds directly to the ordered noisy $p$-values $\tilde{p}^{(0)}_{(1)}, \dots, \tilde{p}^{(0)}_{(m)}$. For instance, when we choose BH-type threshold $\lambda_j = \alpha j/m$,  Algorithm~\ref{alg:SUP-Truncated} rejects the set $\{j: \tilde{p}^{(0)}_{j} \leq \alpha j^{*}/m\}$, where $j^{*} = \sup\{j: \tilde{p}_{(j)}^{(0)} \leq \lambda_{j}, j \leq m\}$.

Within the following Bayesian framework, the first component of power loss can be quantified by analyzing the discrepancy between the distributions of the original $p$-values $p_1,\dots,p_m$ and their privatized counterparts $\tilde{p}^{(0)}_{1}, \dots, \tilde{p}^{(0)}_{m}$.
Let $\theta_1, \dots, \theta_m$ be independent Bernoulli variables, where $\theta_j = 1$ indicates a non-null hypothesis with probability $0<\omega_1<1$ and $\theta_j = 0$ indicates a null hypothesis with probability $\omega_0 = 1 - \omega_1>0$.
We consider each $p$-value is independently generated from the mixture model \citep{efron2001empirical,efron2004large,sun2007oracle} 
\begin{equation}
p_j \mid \theta_j \sim (1-\theta_j) f_{0}(p)+ \theta_j f_{1}(p), \quad 0 \leq p \leq 1,
\end{equation}
where $f_{0}(p)$ and $f_{1}(p)$ are continuous null and non-null densities, respectively. We further denote the cumulative distribution function of
$p_j$ as $F(p) = \omega_0F_0(p) + \omega_1F_1(p)$. Since each original $p$-value is perturbed by independent and identically distributed noise, the distribution of each noise $p$-value $\tilde{p}^{(0)}_j$ can be formulated as 
\begin{equation}
\tilde{p}^{(0)}_j \mid \theta_j \sim \tilde{f}(p)= (1 - \theta_j) \tilde{f}_{0}(p)+ \theta_j \tilde{f}_{1}(p),
\end{equation}
where $\tilde{f}_{0}(p)$ and $\tilde{f}_{1}(p)$ are the null and non-null densities after privatization, respectively. The corresponding cumulative distribution function is $\tilde{F}(p) = \omega_0\tilde{F}_0(p) + \omega_1\tilde{F}_1(p)$. For instance, consider $p_j = \Phi(T_j)$, where $T_j$ follows $N(0,1)$ under the null and $N(a,1)$ under the alternative with $a < 0$. In this case, we have $F_0(p) = p$ and $F_1(p) = \Phi(\Phi^{-1}(p)+a)$. For the noisy $p$-value $\tilde{p}^{(0)}_j$ generated in Algorithm~\ref{alg:SUP-Truncated}, it follows that $\tilde{F}_0(p) = p$ and
$\tilde{F}_1(p) = \Phi\big(\Phi^{-1}(p)+{a}{(1+ 2m'\operatorname{GS}^2_{Q_p}\mu^{-2})^{-1/2}} \big).$

\begin{figure}[htbp]
    \centering
    \includegraphics[height=4cm]{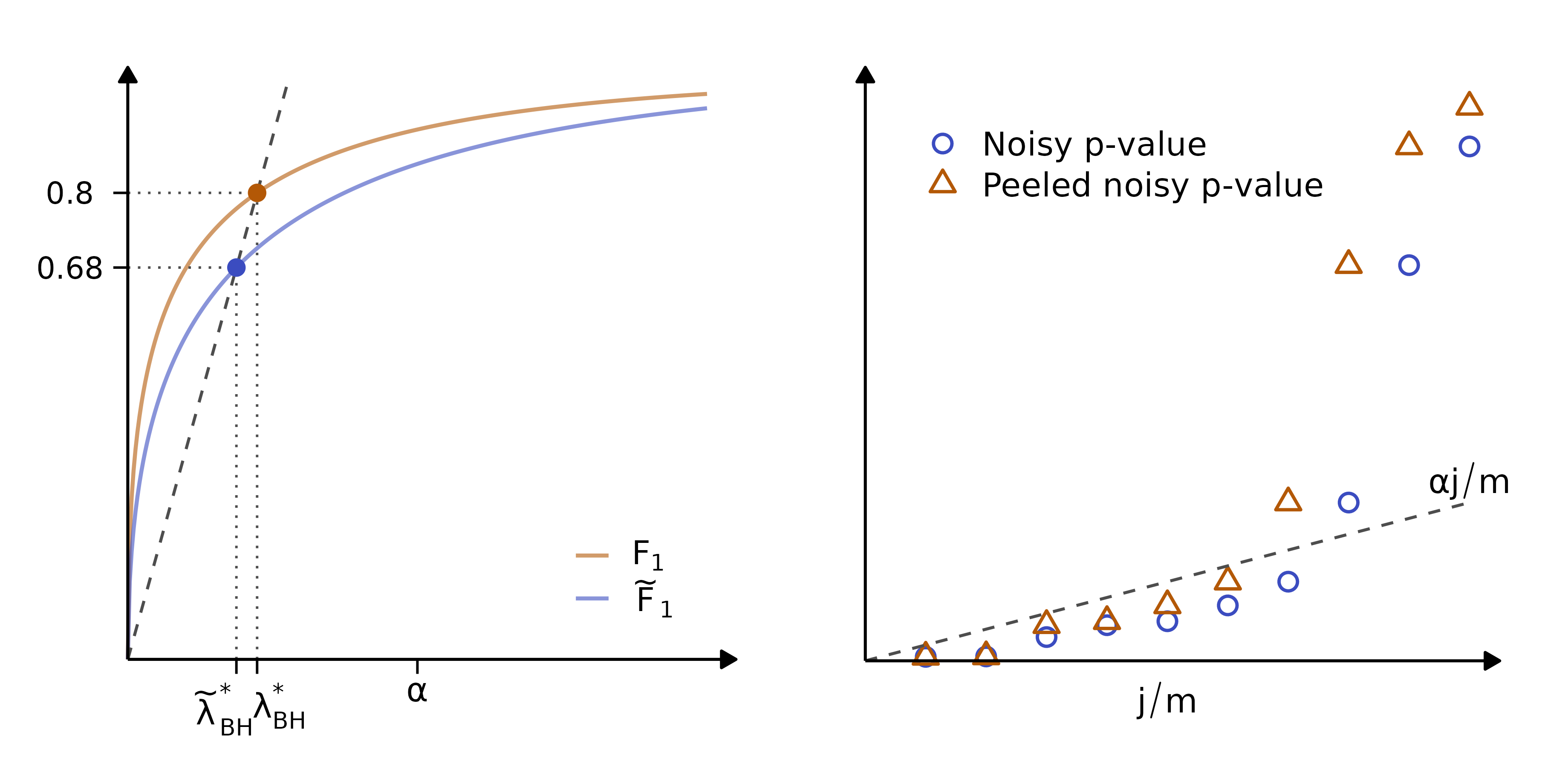}
    \caption{\textit{Left panel: Geometry interpretation of the power loss when the BH threshold is applied to original versus noisy $p$-values. The brown and blue curves display the cumulative distribution functions of the non-null $p_j$ and $\tilde{p}^{(0)}_j$, respectively. Right panel: Noisy $p$-values $\tilde{p}^{(0)}_{(1)}, \dots, \tilde{p}^{(0)}_{(10)}$ (blue dots) and peeled noisy $p$-values $\tilde{p}^{\mathcal{J}}_{(1)}, \dots, \tilde{p}^{\mathcal{J}}_{(10)}$ (brown triangles) compared with the BH threshold. $\alpha=0.2$.}}
    \label{Fig: NoiPower}
\end{figure}

Next, we quantify the power loss of Algorithm \ref{alg:SUP-Truncated} employing BH threshold relative to the non-private BH procedure. Denote $\lambda_{\operatorname{BH}}$ and $\tilde{\lambda}_{\operatorname{BH}}$ as the rejection thresholds determined by the non-private BH procedure and Algorithm~\ref{alg:SUP-Truncated} employing BH threshold. Define the true discovery proportions as
$$
\operatorname{TDP_{\operatorname{BH}}} = \frac{\sum_{j=1}^m \mathbbm{1}(j \in S_1, p_j \leq \lambda_{\operatorname{BH}})}{\sum_{j=1}^m \mathbbm{1}(j \in S_1)}, \quad \widetilde{\operatorname{TDP}}_{\operatorname{BH}} = \frac{\sum_{j=1}^m \mathbbm{1}(j \in S_1, \tilde{p}^{(0)}_j \leq \tilde{\lambda}_{\operatorname{BH}})}{\sum_{j=1}^m \mathbbm{1}(j \in S_1)}.
$$
As illustrated in \citet{genovese2002operating}, the rejection threshold $\lambda_{\operatorname{BH}}$ converges to $\lambda^{*}_{\operatorname{BH}} = \sup\{p:F_1(p) = \beta p\}$ as $m$ approaches infinity, where $\beta = (1- \alpha\omega_0)(\alpha\omega_1)^{-1}$. Analogously, under the noisy scenario, the threshold $\tilde{\lambda}_{\operatorname{BH}}$ converges to $\tilde{\lambda}^{*}_{\operatorname{BH}} = \sup\{p:\tilde{F}_1(p) = \beta p\}$. As depicted in the left panel of Figure~\ref{Fig: NoiPower}, $\beta$ corresponds to the slope of the dashed line that intersects curves $F_1$ and $\tilde{F}_1$. Let $\psi = \min \{\omega_1\tilde{F}_1(\tilde{\lambda}^{*}_{\operatorname{BH}}), \omega_1F_1(\lambda^{*}_{\operatorname{BH}})\}$, $f_{\vartheta} = \sup\{f(p): \lambda^*_{\operatorname{BH}} - \vartheta \leq p \leq \lambda^*_{\operatorname{BH}} + \vartheta\}$, and $\tilde{f}_{\vartheta} = \sup\{\tilde{f}(p): \tilde{\lambda}^*_{\operatorname{BH}} - \vartheta \leq p \leq \tilde{\lambda}^*_{\operatorname{BH}} + \vartheta\}$.  Theorem~\ref{Theo:sup-power-bayes} shows that the difference between $\operatorname{TDP_{\operatorname{BH}}}$ and $\widetilde{\operatorname{TDP}}_{\operatorname{BH}}$ can be approximated by the vertical distance between these two intersection points, which captures the first component of power loss arising from the privacy mechanism. 
\begin{theorem}[Privacy cost]
    \label{Theo:sup-power-bayes}
    Assume that $f_1$ and $\tilde{f}_1$ are strictly decreasing density functions with  $\min\{f_1(0), \tilde{f}_1(0)\}>\beta$. Then, for a small constant $\vartheta$ and any $0<\epsilon\leq \min\{48\psi^{-1}\vartheta \min\{f_{\vartheta}, \tilde{f}_{\vartheta}\},\psi\}$, there exist positive constants $C_1$ and $C_2$ such that
    $$\mathbb{P}\left(\left|\big(\operatorname{TDP_{\operatorname{BH}}} - \widetilde{\operatorname{TDP}}_{\operatorname{BH}}\big)-\big(F_1(\lambda^{*}_{\operatorname{BH}}) - \tilde{F}_1(\tilde{\lambda}^{*}_{\operatorname{BH}})\big)\right| > \epsilon \right) \leq (2m+C_1)e^{-C_2m\epsilon^2}.$$
\end{theorem}
Note that most multiple testing procedures ultimately reject hypotheses with corresponding $p$-values fall below a certain threshold, we can similarly analyze the power loss of  Algorithm~\ref{alg:SUP-Truncated} using other thresholds by examining the discrepancy between distributions $F_1$ and $\tilde{F}_1$.

The second component of power loss stems from the random peeling step and is challenging to quantify precisely due to the complexity inherent in the peeling procedure.
Nevertheless, Theorem~\ref{Theo:sup-power-upperbound} and subsequent analyses establish a connection between the statistical powers of Algorithms~\ref{alg:SUP} and \ref{alg:SUP-Truncated} by examining the property of peeled noisy $p$-values.
\begin{theorem}[Upper bound on power]
    \label{Theo:sup-power-upperbound} 
    If Algorithm~\ref{alg:SUP} and \ref{alg:SUP-Truncated} employ the same input parameters, the power of Algorithm~\ref{alg:SUP} is at most no more than that of Algorithm~\ref{alg:SUP-Truncated}.
\end{theorem}

As illustrated in the right panel of Figure~\ref{Fig: NoiPower}, the underlying principle of Theorem~\ref{Theo:sup-power-upperbound} is that, for $j = 1,\dots,m'$, the $j$-th smallest peeled noisy $p$-value is greater than or equal to the $j$-th smallest noisy $p$-value without peeling, i.e., $\tilde{p}_{(j)}^{(0)} \leq \tilde{p}_{(j)}^{\mathcal{J}}$.
Denote $\mathcal{J}$ and $\mathcal{J}^{\star}$ as the sets of peeled indices and finally rejected indices by Algorithm~\ref{alg:SUP}. Let $\mathcal{J}^{*}$ be the set of finally rejected indices by Algorithm~\ref{alg:SUP-Truncated}. 
Notice that, for any threshold employed in this paper, $j^{\star}$ is a non-increasing function of the peeled noisy $p$-values, it holds that $\mathcal{J}^{\star} \subseteq \mathcal{J}^{*}$.
Furthermore, if $\mathcal{J}^{*} \subseteq \mathcal{J}$, then it follows that $\mathcal{J}^{\star} = \mathcal{J}^{*}$. In this case, the two algorithms produce both the same true discoveries and false discoveries.
Intuitively, when the privacy requirement is not too stringent, a sufficiently large peeling number guarantees that $\mathcal{J}^* \subseteq \mathcal{J}$ with high probability. The experiment results in Figure~\ref{Fig: Inclusion} in Supplementary Material~\ref{sup:numer} confirms that when the peeling number exceeds the true number of signals, the two algorithms exhibit nearly identical power. 

A comparative analysis of statistical power has been conducted between the proposed framework and its non-private counterpart. Next, we demonstrate the superiority of the proposed framework over existing private methods in terms of power. Notice that DP-BH and DP-Bonf methods \citep{dwork2021differentially} ensure $(\varepsilon,\delta)$-DP using Laplace noise and $(\eta, \nu)$-sensitivity (refer to definition \ref{def:Multiplicative Sensitivity} in Supplementary Material~\ref{sec:A}). 
To enable a fair comparison, we extend the proposed SUP framework to ensure \((\varepsilon, \delta)\)-DP. For brevity, detailed Algorithms~\ref{alg:SUP_Laplace}-\ref{alg:SUPBonf_Laplace} are deferred to Supplementary Material~\ref{sup:laplace}. Theorem~\ref{Theo:sup-power} illustrates the superiority of SUP compared to DP-BH and DP-Bonf in terms of power, given identical sensitivity and privacy parameters. 
\begin{theorem}[Superiority in power]
    \label{Theo:sup-power} 
    Given any \( m \) original $p$-values:
    \begin{enumerate}[(1).]
        \item If the parameters satisfy that $\varepsilon^{-1}\eta \sqrt{10 m' \log (1 / \delta)} \leq 1 - (\log \left(6 m' / \alpha\right))^{-1}$, Algorithm~\ref{alg:SUP_Laplace} with the SUP BH threshold is superior to the DP-BH algorithm in terms of power.
        \item If the parameters satisfy that $0.5\varepsilon^{-1}\eta \sqrt{10 m \log (1 / \delta)} \leq 1 - (\log \left(5 m / \alpha\right))^{-1}$, Algorithm~\ref{alg:SUPBonf_Laplace} is superior to the DP-Bonf algorithm in terms of power.
    \end{enumerate}
\end{theorem}
The conditions specified in Theorem~\ref{Theo:sup-power} are confirmed to hold across all simulation settings considered in  \citet{dwork2021differentially}. 
We remark here that the superior performance of our methods arises from incorporating the noise distribution information when generating noisy \(p\)-values, which preserves the super uniform property of the \(p\)-values. This eliminates the need to account for the noise distribution again when constructing thresholds. Thus, our thresholds are insensitive to privacy parameters. In contrast, the methods proposed by \citet{dwork2021differentially} disrupt the super uniform property of the \(p\)-values, necessitating the incorporation of noise distribution information during threshold construction. As a consequence, DP-BH employs the threshold \(\log(\alpha j / m) - \eta \sqrt{10m' \log(1/\delta) \log(6m'/\alpha)}/\varepsilon\) for \(j = 1, \dots, m'\), while DP-Bonf uses the threshold \(\log(\alpha / m) - \eta \sqrt{10m \log(1/\delta) \log(5m/\alpha)}/(2\varepsilon)\). These thresholds are highly sensitive to privacy parameters, becoming increasingly conservative as privacy guarantees are strengthened, which may reduce statistical power.

\vspace{-0.9ex}
\section{Simulation}\label{sec:simu}
In this section, we present a series of simulations to assess the effectiveness of the proposed framework in controlling FDR and FWER, as well as its statistical power. 

The simulation parameters are set as follows unless otherwise specified. The total number of hypotheses is fixed at \( m = 20,000 \).
Let $T = \left(T_1, \dots, T_m\right)$ be a vector following multivariate normal distribution $N(0_m, \Sigma)$, where the covariance matrix $\Sigma = (\sigma_{ij})_{m \times m}$ has diagonal elements set to 1. For the independent case, the off-diagonal elements of $\Sigma$ are set to 0. For the dependent case, $\Sigma$ is structured as a blockwise diagonal matrix consisting of 100 submatrices, with $\sigma_{ij}=0.6$ for $i\neq j$ within each submatrix, and the elements outside these submatrices are set to 0. We define the null hypotheses as \(\theta_j \leq 0\) and the non-null hypotheses as \(\theta_j > 0\), and each $p$-value is generated as $p_j=\Phi\left(T_j-\theta_j\right)$.
We randomly choose \( m_1 = 100 \) non-null hypotheses to have signal coefficients of $\theta_j = 4$. For the remaining \( m_0 = m - m_1 \) null hypotheses, we explore two scenarios. In the uniform setting, all null $\theta_j$ are set to 0. In the conservative setting, where some $p_j$ satisfy $\mathbb{P}(p_j \leq t) < t$ for $t \in (0,1)$, we randomly choose $0.6m_0$ null $p$-values to have signal coefficients of $\theta_j = 0$ and the remaining $0.4m_0$ null $p$-values are assigned signal coefficients drawn from $U(-0.3,0)$. This conservative setting satisfies both the super uniform condition and the mirror conservative condition, often encountered in one-sided hypothesis testing. 

For FDR control, we compare Algorithm~\ref{alg:SUP} with the SUP BH threshold, referred to as SUP-BH, and Algorithm~\ref{alg:SUP} employing the SUP Benjamini-Yekutieli threshold, referred to as SUP-BY, as well as Algorithm~\ref{alg:AdaSUP} with the Adaptive SUP BH threshold, referred to as ASUP-BH, against three baseline methods: the non-private BH method \citep{benjamini1995controlling}, the DP-BH method \citep{dwork2021differentially}, and the DP-AdaPT method \citep{xia2023adaptive}. 
In terms of FWER control, we compare Algorithm~\ref{alg:SUP} with the SUP Bonferroni threshold, referred to as SUP-Bonf, and the SUP Holm threshold, referred to as SUP-Holm, as well as Algorithm~\ref{alg:AdaSUP} with the Adaptive SUP Bonferroni threshold, referred to as ASUP-Bonf, against two baseline methods: the non-private Bonferroni method and the DP-Bonf method \citep{dwork2021differentially}.
We remark that the DP-BH and DP-Bonf methods employ Laplace noise to ensure \((\varepsilon, \delta)\)-DP, whereas the DP-AdaPT method uses Gaussian noise to ensure \(\mu\)-GDP. For fairness, we utilize Gaussian noise for SUP in this section and conduct additional simulations with Laplace noise, as detailed in Supplementary Material~\ref{sup:laplace}.
For a given Type-I error level $\alpha$, denote 
$$\widehat{\operatorname{FDR}}_{\alpha}= \frac{|\mathcal{R}_\alpha \cap S_0|}{|\mathcal{R}_\alpha| \vee 1}, \;\; \widehat{\operatorname{FWER}}_\alpha = \mathbbm{1}(|\mathcal{R}_\alpha \cap S_0| > 0),  \;\; \widehat{\operatorname{Power}}_{\alpha}= \frac{|\mathcal{R}_\alpha \cap S_1|}{m_1 \vee 1}, $$
where  $\mathcal{R}_\alpha$ represents the index set of rejected hypotheses. In this section, we set $\alpha = 0.1$.
The empirical FDR, FWER, and power are respectively calculated by averaging $\widehat{\operatorname{FDR}}_{\alpha}$, $\widehat{\operatorname{FWER}}_\alpha$, and $\widehat{\operatorname{Power}}_{\alpha}$ over 200 replications.

To ensure consistency and fairness, our parameter settings align with those in \citet{dwork2021differentially, xia2023adaptive}. 
For the DP-BH and DP-Bonf methods, the sensitivity parameters are set to \( \eta = 10^{-4} \) and \( \nu = 0.5 \alpha / m \), respectively. The privacy parameters are set to $\varepsilon = 0.5$ and $\delta = 0.001$. Both the DP-AdaPT algorithm and our proposed frameworks adopt the privacy parameter $\mu = 4 \varepsilon / \sqrt{10 \log (1 / \delta)}$, the transformation $Q(\cdot) = \Phi^{-1}(\cdot)$, and the global sensitivity $\operatorname{GS}_{Q_p} = \eta$, following \citet{xia2023adaptive}.
We set the peeling number $m' = 200$ for the DP-BH, DP-AdaPT, SUP-BH, SUP-BY, SUP-Bonf, and SUP-Holm procedures. For ASUP-BH and ASUP-Bonf procedures, the minimum peeling number $\tilde{m}$ is set to 100.

\begin{figure}[htbp]
    \centering
    \includegraphics[height=5.8cm]{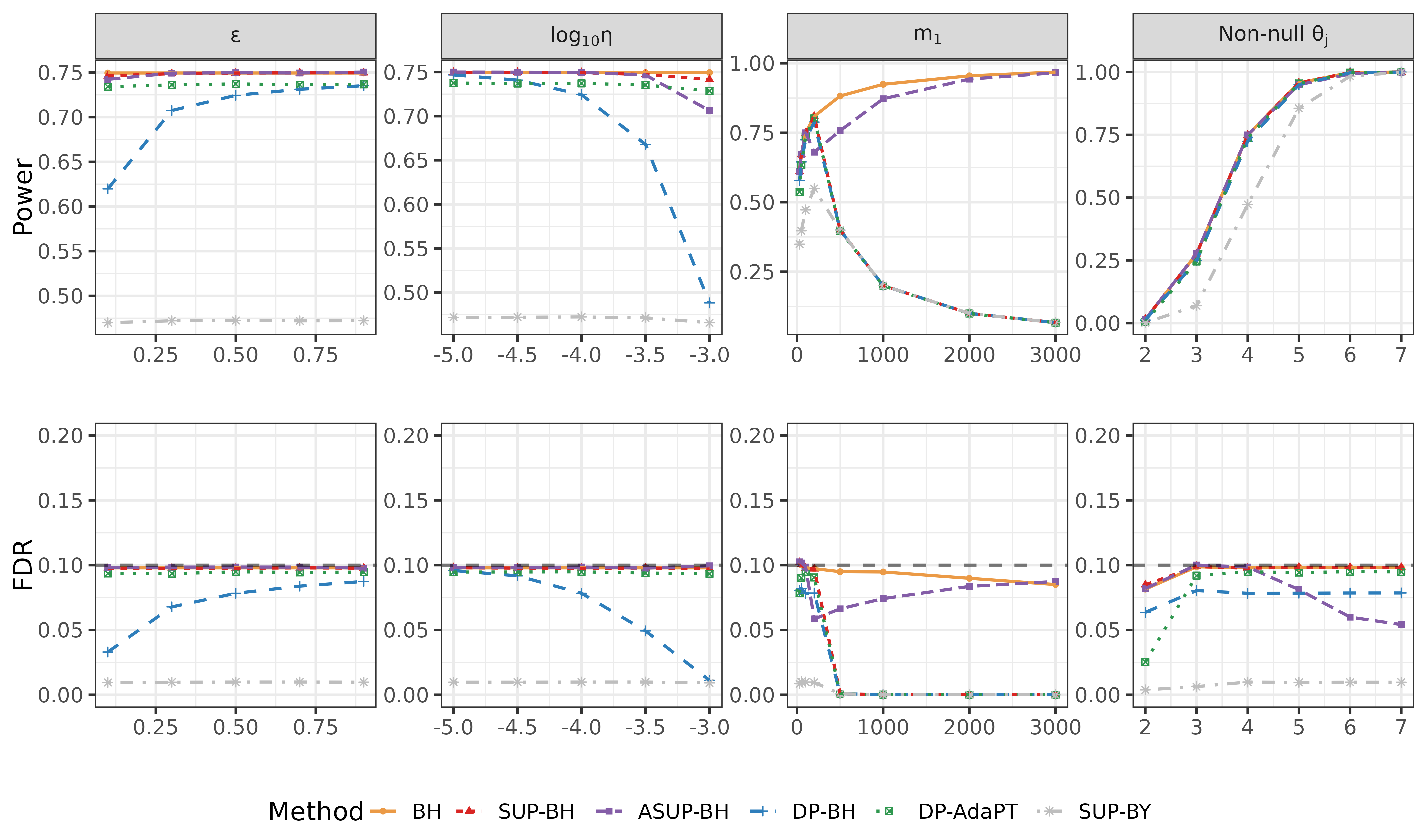}
    \caption{\textit{The FDRs and powers for BH, DP-BH, DP-AdaPT, SUP-BH, and ASUP-BH under varying parameters with independent and uniform $p$-values.
    }}
    \label{Fig: Uniform}
\end{figure}

\begin{figure}[htbp]
    \centering
    \includegraphics[height=5.8cm]{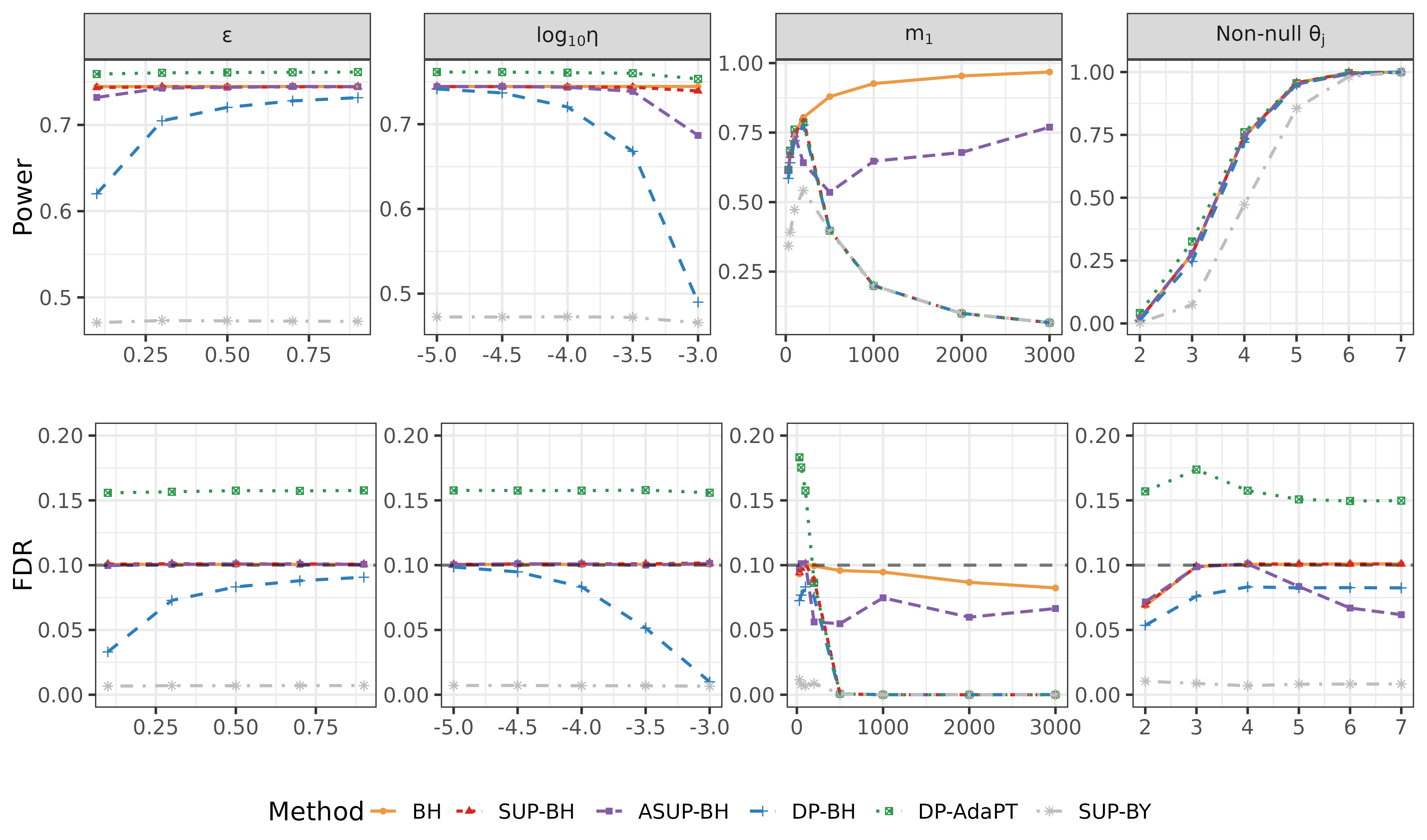}
    \caption{\textit{The FDRs and powers for BH, DP-BH, DP-AdaPT, SUP-BH, and ASUP-BH under varying parameters with dependent and uniform $p$-values.
    }}
    \label{Fig: Dependence}
\end{figure}
Figures~\ref{Fig: Uniform}-\ref{Fig: Dependence} and \ref{Fig: MixGau} in Supplementary Material~\ref{sup:numer} present the results of FDR and power against varying privacy parameter \(\varepsilon\), sensitivity parameter \(\eta\), number of non-null hypotheses \(m_1\), and \(\theta_j\) for non-null hypotheses.
The result under uniform and independent setting is displayed in Figure~\ref{Fig: Uniform}. It shows that all methods consistently control the FDR at the target level. However, the DP-BH's FDR is sensitive to \(\varepsilon\) and \(\eta\) and becomes conservative when \(\varepsilon\) is small and \(\eta\) is large. Both SUP-BH and ASUP-BH showcase superior power, with ASUP-BH achieving greater advantages as the number of non-null hypotheses \(m_1\) exceeds the peeling number. SUP-BH exhibits only mild power loss while significantly enhancing privacy compared to non-private BH. Figure~\ref{Fig: Dependence} reveals that DP-AdaPT fails to control the FDR under dependence, whereas the other methods consistently maintain FDR control. For the conservative setting, Figure~\ref{Fig: MixGau} shows that all methods conservatively maintain FDR control at the desired level. The results of power are similar to those in Figure~\ref{Fig: Uniform}. Please see Supplementary Material~\ref{sup:numer} for details on FWER.

\section{Real data analysis}\label{sec:real}
In this section, we apply our framework to a myeloma dataset originally collected by \citet{tian2003Erm} and later utilized for FWER and FDR analysis by \citet{zehetmayer2008optimized} and \citet{fu2022heteroscedasticity}.
It comprises gene expression data from 173 patients diagnosed with multiple myeloma, classified into two groups based on MRI findings: 137 patients with focal bone lesions detected by MRI, and 36 patients without such detectable lesions. The primary objective of this study is to identify molecular determinants of osteolytic lesions by testing differences in gene expression patterns across 12,625 genes between the two patient groups. For each gene, the Wilcoxon rank-sum test is applied to compute the $p$-value. This dataset is publicly available at \href{http://www.ncbi.nlm.nih.gov/geo/query/acc.cgi?acc=GSE755}{\textit{http://www.ncbi.nlm.nih.gov/geo/query/acc.cgi?acc=GSE755}}. 

In this experiment, we evaluate Type-I error level $\alpha$ ranging from 0.02 to 0.2 at equal intervals. The sensitivity \(\eta\) is chosen as \(3 \times 10^{-4}\), and the truncation threshold \(\nu\) is set to \(0.3 \alpha / m\) for DP-BH and DP-Bonf. The privacy parameters are fixed at \(\varepsilon = 0.5\) and \(\delta = 0.001\). Both DP-AdaPT and our proposed methods utilize the privacy parameter \(\mu = 4 \varepsilon / \sqrt{10 \log (1 / \delta)}\), the Gaussian quantile transformation \(Q(\cdot) = \Phi^{-1}(\cdot)\), and global sensitivity $\operatorname{GS}_{Q_p} = \eta$, in line with the setting suggested by \citet{xia2023adaptive}.
We examine various peeling numbers, \(m' \in \{100, 500, 1500, 4000\}\), for DP-BH, DP-AdaPT, SUP-BH, SUP-BY, SUP-Bonf, SUP-Holm  methods. For ASUP-BH and ASUP-Bonf, the minimum peeling number $\tilde{m}$ is set to 100. Considering the inherent randomness of the added noise, we repeat the experiment 200 times and compute the average number of rejections.

\begin{figure}[htbp]
    \centering
    \includegraphics[height=3.5cm]{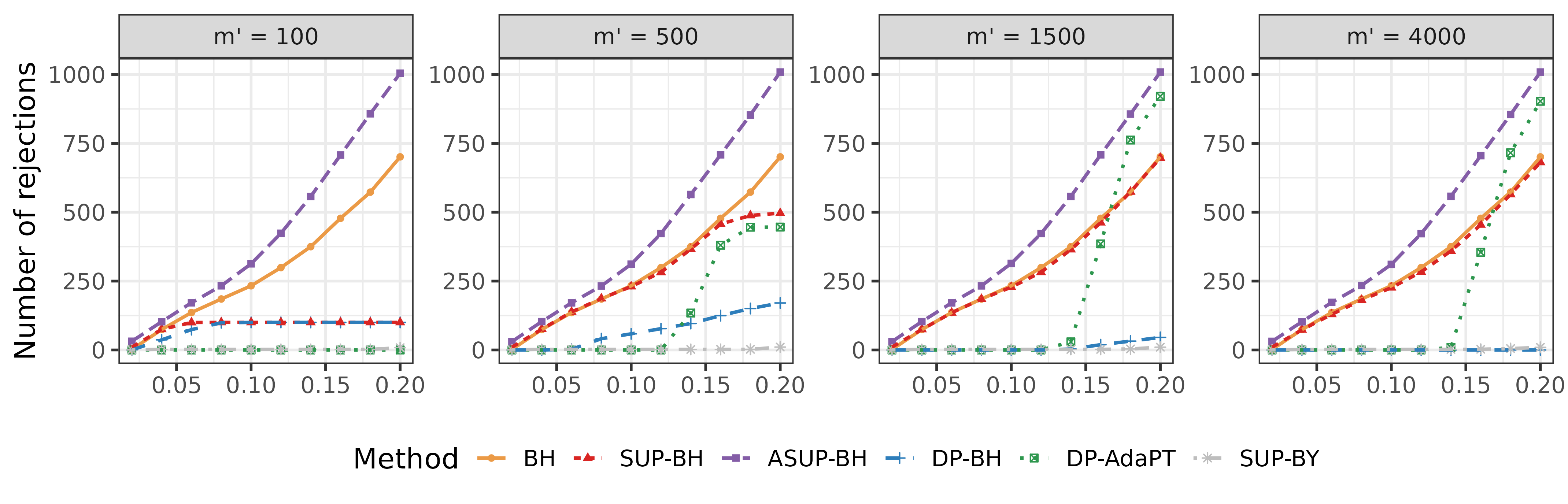}
    \caption{\textit{The number of rejections for BH, DP-BH, DP-AdaPT, SUP-BH, SUP-BY, and ASUP-BH across varying FDR levels and peeling numbers.
    }}
    \label{Fig: tian_FDR}
\end{figure}

Figure~\ref{Fig: tian_FDR} compares the number of rejections across various methods, highlighting the impact of the peeling number $m'$ on DP-BH, DP-AdaPT, and SUP-BH. Specifically, DP-BH becomes overly conservative for large $m'$, resulting in very few rejections, whereas SUP-BH closely matches the performance of the non-private BH method. DP-AdaPT initially shows an increase and then a decline in rejections as $m'$ grows, making fewer rejections at low FDR levels. Among these methods, ASUP-BH achieves the highest number of rejections. To assess each method’s stability, we augment the original $p$-values with synthetic ones and analyze the combined set. Details are provided in Supplementary Material~\ref{sup:numer}, which also contains the corresponding FWER results.

\vspace{-0.9ex}
\section{Discussion}\label{sec:dis}
This paper introduces a private multiple testing framework designed to manage both FWER and FDR control, irrespective of the dependence among $p$-values. We propose a novel transformation of $p$-values, a reversed peeling algorithm, a range of rejection thresholds, and adaptive techniques to determine the peeling number and adjust thresholds. Through theoretical analysis and empirical investigations, we demonstrate that our methods provide robust Type-I errors control and surpass existing methods in terms of power.

To conclude, this work highlights some valuable questions and opens promising avenues for future research. First, the proposed framework could be generalized to incorporate heterogeneous covariate information, combining it with existing structure-adaptive multiple testing methods that rely on the super uniform condition \citep{li2019multiple, ignatiadis2021covariate}. 
Second, variable selection can often be framed as multiple testing, where constructing \(p\)-values can be particularly challenging in high-dimensional settings. Methods that utilize symmetric statistics to control FDR are promising \citep{candes2018panning,du2021false,dai2022false,wang2025adaptive}. It is worth to explore how to provide privacy protection while maintaining effectiveness in FDR control for these methods.  Last but not least, since online multiple testing is widely used by major technology companies and in platform trials, and privacy protection is even more challenging in the online setting, we are committed to developing an online adaptation of our framework.





\setstretch{1.23}
\bibliographystyle{abbrvnat}
\bibliography{reference}

\begin{thebibliography}{40}
\providecommand{\natexlab}[1]{#1}
\providecommand{\url}[1]{\texttt{#1}}
\expandafter\ifx\csname urlstyle\endcsname\relax
  \providecommand{\doi}[1]{doi: #1}\else
  \providecommand{\doi}{doi: \begingroup \urlstyle{rm}\Url}\fi

\bibitem[Benjamini and Hochberg(1995)]{benjamini1995controlling}
Y.~Benjamini and Y.~Hochberg.
\newblock Controlling the false discovery rate: a practical and powerful approach to multiple testing.
\newblock \emph{Journal of the Royal Statistical Society Series B: Statistical Methodology}, 57\penalty0 (1):\penalty0 289--300, 1995.

\bibitem[Benjamini and Yekutieli(2001)]{benjamini2001control}
Y.~Benjamini and D.~Yekutieli.
\newblock The control of the false discovery rate in multiple testing under dependency.
\newblock \emph{The Annals of Statistics}, 29\penalty0 (4):\penalty0 1165 -- 1188, 2001.

\bibitem[Bonferroni(1936)]{bonferroni1936teoria}
C.~Bonferroni.
\newblock Teoria statistica delle classi e calcolo delle probabilita.
\newblock \emph{Pubblicazioni del R Istituto Superiore di Scienze Economiche e Commericiali di Firenze}, 8:\penalty0 3--62, 1936.

\bibitem[Candes et~al.(2018)Candes, Fan, Janson, and Lv]{candes2018panning}
E.~Candes, Y.~Fan, L.~Janson, and J.~Lv.
\newblock Panning for gold:‘model-x’knockoffs for high dimensional controlled variable selection.
\newblock \emph{Journal of the Royal Statistical Society Series B: Statistical Methodology}, 80\penalty0 (3):\penalty0 551--577, 2018.

\bibitem[Dai et~al.(2023)Dai, Lin, Xing, and Liu]{dai2022false}
C.~Dai, B.~Lin, X.~Xing, and J.~S. Liu.
\newblock False discovery rate control via data splitting.
\newblock \emph{Journal of the American Statistical Association}, 118\penalty0 (544):\penalty0 2503--2520, 2023.

\bibitem[Dong et~al.(2022)Dong, Roth, and Su]{dong2022gaussian}
J.~Dong, A.~Roth, and W.~J. Su.
\newblock Gaussian differential privacy.
\newblock \emph{Journal of the Royal Statistical Society Series B: Statistical Methodology}, 84\penalty0 (1):\penalty0 3--37, 2022.

\bibitem[Du et~al.(2023)Du, Guo, Sun, and Zou]{du2021false}
L.~Du, X.~Guo, W.~Sun, and C.~Zou.
\newblock False discovery rate control under general dependence by symmetrized data aggregation.
\newblock \emph{Journal of the American Statistical Association}, 118\penalty0 (541):\penalty0 607--621, 2023.

\bibitem[Duggal et~al.(2008)Duggal, Gillanders, Holmes, and Bailey-Wilson]{duggal2008establishing}
P.~Duggal, E.~M. Gillanders, T.~N. Holmes, and J.~E. Bailey-Wilson.
\newblock Establishing an adjusted p-value threshold to control the family-wide type 1 error in genome wide association studies.
\newblock \emph{BMC Genomics}, 9:\penalty0 1--8, 2008.

\bibitem[Dwork et~al.(2006)Dwork, McSherry, Nissim, and Smith]{dwork2006calibrating}
C.~Dwork, F.~McSherry, K.~Nissim, and A.~Smith.
\newblock Calibrating noise to sensitivity in private data analysis.
\newblock In \emph{Theory of Cryptography: Third Theory of Cryptography Conference, TCC 2006, New York, NY, USA, March 4-7, 2006. Proceedings 3}, pages 265--284. Springer, 2006.

\bibitem[Dwork et~al.(2014)Dwork, Roth, et~al.]{dwork2014algorithmic}
C.~Dwork, A.~Roth, et~al.
\newblock The algorithmic foundations of differential privacy.
\newblock \emph{Foundations and Trends{\textregistered} in Theoretical Computer Science}, 9\penalty0 (3--4):\penalty0 211--407, 2014.

\bibitem[Dwork et~al.(2021)Dwork, Su, and Zhang]{dwork2021differentially}
C.~Dwork, W.~Su, and L.~Zhang.
\newblock Differentially private false discovery rate control.
\newblock \emph{Journal of Privacy and Confidentiality}, 11\penalty0 (2), 2021.

\bibitem[Efron(2004)]{efron2004large}
B.~Efron.
\newblock Large-scale simultaneous hypothesis testing: the choice of a null hypothesis.
\newblock \emph{Journal of the American Statistical Association}, 99\penalty0 (465):\penalty0 96--104, 2004.

\bibitem[Efron et~al.(2001)Efron, Tibshirani, Storey, and Tusher]{efron2001empirical}
B.~Efron, R.~Tibshirani, J.~D. Storey, and V.~Tusher.
\newblock Empirical bayes analysis of a microarray experiment.
\newblock \emph{Journal of the American Statistical Association}, 96\penalty0 (456):\penalty0 1151--1160, 2001.

\bibitem[Finner et~al.(2009)Finner, Dickhaus, and Roters]{finner2009false}
H.~Finner, T.~Dickhaus, and M.~Roters.
\newblock On the false discovery rate and an asymptotically optimal rejection curve.
\newblock \emph{The Annals of Statistics}, 37\penalty0 (2):\penalty0 596 -- 618, 2009.

\bibitem[Fu et~al.(2022)Fu, Gang, James, and Sun]{fu2022heteroscedasticity}
L.~Fu, B.~Gang, G.~M. James, and W.~Sun.
\newblock Heteroscedasticity-adjusted ranking and thresholding for large-scale multiple testing.
\newblock \emph{Journal of the American Statistical Association}, 117\penalty0 (538):\penalty0 1028--1040, 2022.

\bibitem[Genovese and Wasserman(2002)]{genovese2002operating}
C.~Genovese and L.~Wasserman.
\newblock Operating characteristics and extensions of the false discovery rate procedure.
\newblock \emph{Journal of the Royal Statistical Society Series B: Statistical Methodology}, 64\penalty0 (3):\penalty0 499--517, 2002.

\bibitem[Greenstreet et~al.(2021)Greenstreet, Jaki, Bedding, Harbron, and Mozgunov]{greenstreet2021multi}
P.~Greenstreet, T.~Jaki, A.~Bedding, C.~Harbron, and P.~Mozgunov.
\newblock A multi-arm multistage platform design that allows preplanned addition of arms while still controlling the family-wise error.
\newblock \emph{Statistics in Medicine}, 43\penalty0 (19):\penalty0 3613--3632, 2021.

\bibitem[Hochberg(1988)]{hochberg1988sharper}
Y.~Hochberg.
\newblock A sharper bonferroni procedure for multiple tests of significance.
\newblock \emph{Biometrika}, 75\penalty0 (4):\penalty0 800--802, 1988.

\bibitem[Holm(1979)]{holm1979simple}
S.~Holm.
\newblock A simple sequentially rejective multiple test procedure.
\newblock \emph{Scandinavian Journal of Statistics}, 6\penalty0 (2):\penalty0 65--70, 1979.

\bibitem[Homer et~al.(2008)Homer, Szelinger, Redman, Duggan, Tembe, Muehling, Pearson, Stephan, Nelson, and Craig]{homer2008resolving}
N.~Homer, S.~Szelinger, M.~Redman, D.~Duggan, W.~Tembe, J.~Muehling, J.~V. Pearson, D.~A. Stephan, S.~F. Nelson, and D.~W. Craig.
\newblock Resolving individuals contributing trace amounts of dna to highly complex mixtures using high-density snp genotyping microarrays.
\newblock \emph{PLoS Genetics}, 4\penalty0 (8):\penalty0 e1000167, 2008.

\bibitem[Ignatiadis and Huber(2021)]{ignatiadis2021covariate}
N.~Ignatiadis and W.~Huber.
\newblock Covariate powered cross-weighted multiple testing.
\newblock \emph{Journal of the Royal Statistical Society Series B: Statistical Methodology}, 83\penalty0 (4):\penalty0 720--751, 2021.

\bibitem[Kairouz et~al.(2017)Kairouz, Oh, and Viswanath]{Kairouz2017the}
P.~Kairouz, S.~Oh, and P.~Viswanath.
\newblock The composition theorem for differential privacy.
\newblock \emph{IEEE Transactions on Information Theory}, 63\penalty0 (6):\penalty0 4037--4049, 2017.

\bibitem[Kaplanis et~al.(2020)Kaplanis, Samocha, Wiel, Zhang, Arvai, Eberhardt, Gallone, Lelieveld, Martin, McRae, et~al.]{kaplanis2020evidence}
J.~Kaplanis, K.~E. Samocha, L.~Wiel, Z.~Zhang, K.~J. Arvai, R.~Y. Eberhardt, G.~Gallone, S.~H. Lelieveld, H.~C. Martin, J.~F. McRae, et~al.
\newblock Evidence for 28 genetic disorders discovered by combining healthcare and research data.
\newblock \emph{Nature}, 586\penalty0 (7831):\penalty0 757--762, 2020.

\bibitem[Korthauer et~al.(2019)Korthauer, Kimes, Duvallet, Reyes, Subramanian, Teng, Shukla, Alm, and Hicks]{korthauer2019practical}
K.~Korthauer, P.~K. Kimes, C.~Duvallet, A.~Reyes, A.~Subramanian, M.~Teng, C.~Shukla, E.~J. Alm, and S.~C. Hicks.
\newblock A practical guide to methods controlling false discoveries in computational biology.
\newblock \emph{Genome Biology}, 20\penalty0 (118):\penalty0 1--21, 2019.

\bibitem[Lei and Fithian(2018)]{lei2018adapt}
L.~Lei and W.~Fithian.
\newblock Ada{PT}: an interactive procedure for multiple testing with side information.
\newblock \emph{Journal of the Royal Statistical Society Series B: Statistical Methodology}, 80\penalty0 (4):\penalty0 649--679, 2018.

\bibitem[Lei et~al.(2020)Lei, Ramdas, and Fithian]{lei2020general}
L.~Lei, A.~Ramdas, and W.~Fithian.
\newblock A general interactive framework for false discovery rate control under structural constraints.
\newblock \emph{Biometrika}, 108\penalty0 (2):\penalty0 253--267, July 2020.

\bibitem[Leung and Sun(2022)]{leung2022zap}
D.~Leung and W.~Sun.
\newblock Zap: z-value adaptive procedures for false discovery rate control with side information.
\newblock \emph{Journal of the Royal Statistical Society Series B: Statistical Methodology}, 84\penalty0 (5):\penalty0 1886--1946, 2022.

\bibitem[Li and Barber(2019)]{li2019multiple}
A.~Li and R.~F. Barber.
\newblock Multiple testing with the structure-adaptive benjamini--hochberg algorithm.
\newblock \emph{Journal of the Royal Statistical Society Series B: Statistical Methodology}, 81\penalty0 (1):\penalty0 45--74, 2019.

\bibitem[Rashkin et~al.(2020)Rashkin, Graff, Kachuri, Thai, Alexeeff, Blatchins, Cavazos, Corley, Emami, Hoffman, et~al.]{rashkin2020pan}
S.~R. Rashkin, R.~E. Graff, L.~Kachuri, K.~K. Thai, S.~E. Alexeeff, M.~A. Blatchins, T.~B. Cavazos, D.~A. Corley, N.~C. Emami, J.~D. Hoffman, et~al.
\newblock Pan-cancer study detects genetic risk variants and shared genetic basis in two large cohorts.
\newblock \emph{Nature Communications}, 11\penalty0 (1):\penalty0 4423, 2020.

\bibitem[Romano and Wolf(2005)]{romano2005exact}
J.~P. Romano and M.~Wolf.
\newblock Exact and approximate stepdown methods for multiple hypothesis testing.
\newblock \emph{Journal of the American Statistical Association}, 100\penalty0 (469):\penalty0 94--108, 2005.

\bibitem[Shi and Wu(2017)]{shi2017overview}
X.~Shi and X.~Wu.
\newblock An overview of human genetic privacy.
\newblock \emph{Annals of the New York Academy of Sciences}, 1387\penalty0 (1):\penalty0 61--72, 2017.

\bibitem[Storey(2002)]{storey2002direct}
J.~D. Storey.
\newblock A direct approach to false discovery rates.
\newblock \emph{Journal of the Royal Statistical Society Series B: Statistical Methodology}, 64\penalty0 (3):\penalty0 479--498, 2002.

\bibitem[Sun and Cai(2007)]{sun2007oracle}
W.~Sun and T.~T. Cai.
\newblock Oracle and adaptive compound decision rules for false discovery rate control.
\newblock \emph{Journal of the American Statistical Association}, 102\penalty0 (479):\penalty0 901--912, 2007.

\bibitem[Tian et~al.(2003)Tian, Zhan, Walker, Rasmussen, Ma, Barlogie, and Shaughnessy]{tian2003Erm}
E.~Tian, F.~Zhan, R.~Walker, E.~Rasmussen, Y.~Ma, B.~Barlogie, and J.~D. Shaughnessy.
\newblock The role of the wnt-signaling antagonist dkk1 in the development of osteolytic lesions in multiple myeloma.
\newblock \emph{New England Journal of Medicine}, 349\penalty0 (26):\penalty0 2483--2494, 2003.

\bibitem[Tian et~al.(2021)Tian, Liang, and Li]{tian2021powerful}
Z.~Tian, K.~Liang, and P.~Li.
\newblock A powerful procedure that controls the false discovery rate with directional information.
\newblock \emph{Biometrics}, 77\penalty0 (1):\penalty0 212--222, 2021.

\bibitem[Wang et~al.(2025)Wang, Chen, Han, Xu, and Kong]{wang2025adaptive}
K.~Wang, Y.~Chen, Y.~Han, W.~Xu, and L.~Kong.
\newblock Adaptive selection for false discovery rate control leveraging symmetry.
\newblock \emph{Journal of the American Statistical Association}, \penalty0 (just-accepted):\penalty0 1--21, 2025.

\bibitem[Wingo et~al.(2021)Wingo, Liu, Gerasimov, Gockley, Logsdon, Duong, Dammer, Robins, Beach, Reiman, et~al.]{wingo2021integrating}
A.~P. Wingo, Y.~Liu, E.~S. Gerasimov, J.~Gockley, B.~A. Logsdon, D.~M. Duong, E.~B. Dammer, C.~Robins, T.~G. Beach, E.~M. Reiman, et~al.
\newblock Integrating human brain proteomes with genome-wide association data implicates new proteins in alzheimer’s disease pathogenesis.
\newblock \emph{Nature Genetics}, 53\penalty0 (2):\penalty0 143--146, 2021.

\bibitem[Xia and Cai(2023)]{xia2023adaptive}
X.~Xia and Z.~Cai.
\newblock Adaptive false discovery rate control with privacy guarantee.
\newblock \emph{Journal of Machine Learning Research}, 24\penalty0 (252):\penalty0 1--35, 2023.

\bibitem[Zehetmayer et~al.(2008)Zehetmayer, Bauer, and Posch]{zehetmayer2008optimized}
S.~Zehetmayer, P.~Bauer, and M.~Posch.
\newblock Optimized multi-stage designs controlling the false discovery or the family-wise error rate.
\newblock \emph{Statistics in Medicine}, 27\penalty0 (21):\penalty0 4145--4160, 2008.

\bibitem[Zhu and Guo(2020)]{zhu2020family}
Y.~Zhu and W.~Guo.
\newblock Family-wise error rate controlling procedures for discrete data.
\newblock \emph{Statistics in Biopharmaceutical Research}, 12\penalty0 (1):\penalty0 117--128, 2020.

\end{thebibliography}

\end{document}